\documentclass[11pt]{article}
\usepackage{amsfonts}
\usepackage{amssymb}
\usepackage{amsmath,amssymb}
\usepackage{bbm}
\usepackage{tikz}

\def\slasha#1{\setbox0=\hbox{$#1$}#1\hskip-\wd0\hbox to\wd0{\hss\sl/\/\hss}}

\def\periodb#1{\setbox0=\hbox{$#1$}#1\hskip-\wd0\hbox to\wd0{-}}

\def\sfrac#1#2{{\textstyle\frac{#1}{#2}}}
\newcommand{\unit}{\mathbbm{1}}   

\newcommand{\CJ}{\mathcal{J}} 
 
\newcommand{\CE}{\mathcal{E}}    
\newcommand{\CL}{\mathcal{L}}    
\newcommand{\CF}{\mathcal{F}}   

\newcommand{\CH}{\mathcal{H}}

\newcommand{\CV}{\mathcal{V}} 
   
\newcommand{\R}{\mathbb{R}}     
\newcommand{\C}{\mathbb{C}}     

\newcommand{\im}{\mathrm{i}} 
 
\newcommand{\dd}{\mathrm{d}}     
\newcommand{\dpar}{\partial}     
\newcommand{\diag}{{\mathrm{diag}}}

\newcommand{\+}{\dagger}

\newcommand{\AdS}{\mathrm{AdS}} 
\newcommand{\sU}{\mathrm{U}}     
\newcommand{\sSU}{\mathrm{SU}}     
     
\newcommand{\sSO}{\mathrm{SO}}     
\newcommand{\sGL}{\mathrm{GL}}

\newcommand{\rS}{\mathrm{S}} 

\newcommand{\al}{{{\alpha}}}

\newcommand{\vk}{{{\varkappa}}}
\newcommand{\veps}{{\varepsilon}} 
\newcommand{\vph}{{{\varphi}}} 
\newcommand{\vrho}{{\varrho}} 
\newcommand{\vsig}{{\varsigma}}

\newcommand{\xh}{\hat{x}} 
\newcommand{\ph}{\hat{p}}

\newcommand{\ab}{{\bar{a}}}
\newcommand{\bb}{{\bar{b}}}
\newcommand{\mb}{{\bar{\mu}}}
\newcommand{\nb}{{\bar{\nu}}}

\newcommand{\qv}{{q_{\sf{v}}}}
\newcommand{\Av}{{A_{\sf{vac}}}}
\newcommand{\Fv}{{F_{\sf{vac}}}}

\newcommand{\und}{\quad\mbox{and}\quad}
\newcommand{\with}{\quad\mbox{with}\quad}
\newcommand{\for}{\quad\mbox{for}\quad}

\makeatletter

\@addtoreset{equation}{section}
\makeatother

\begin{document}
\begin{titlepage}
\setcounter{page}{0}
.
\vskip 15mm
\begin{center}
{\LARGE \bf Supersymmetric Klein-Gordon and Dirac oscillators}\\
\vskip 1.5cm
{\Large Alexander D. Popov}
\vskip 1cm
{\em Institut f\"{u}r Theoretische Physik,
Leibniz Universit\"{a}t Hannover\\
Appelstra{\ss}e 2, 30167 Hannover, Germany}\\
{Email: alexander.popov@itp.uni-hannover.de}
\vskip 1.1cm
\end{center}
\begin{center}
{\bf Abstract}
\end{center}

\noindent 
We have recently shown that the space of initial data (covariant phase space) of the relativistic oscillator in Minkowski space $\R^{3,1}$ is a homogeneous K\"ahler-Einstein manifold $Z_6=\AdS_7/\sU(1) =\sU(3,1)/\sU(3)\times \sU(1)$. It was also shown that the energy eigenstates of the quantum relativistic oscillator form a direct sum of two weighted Bergman spaces of holomorphic (particles) and antiholomorphic (antiparticles) square-integrable functions on the covariant phase space $Z_6$ of the classical oscillator. Here we show that the covariant phase space of the supersymmetric version of the relativistic oscillator (oscillating spinning particle) is the odd tangent bundle of the space $Z_6$. Quantizing this model yields a Dirac oscillator equation on the phase space whose solution space is a direct sum of two spinor spaces parametrized by holomorphic and antiholomorphic functions on the odd tangent bundle of $Z_6$. After expanding the general solution in Grassmann variables, we obtain components of the spinor field that are holomorphic and antiholomorphic functions from Bergman spaces on $Z_6$ with different weight functions. Thus, the supersymmetric model under consideration is exactly solvable, Lorentz covariant and unitary.

\end{titlepage}

\section{Introduction and summary}

\noindent {\bf Free particles}.  A classical {\it non-relativistic} spinless particle of mass $m$ is defined as a point in the phase space $T^*\R^3=\R^3\times\R^3$ parametrized by the coordinates $x^a\in\R^3$ and momenta $p_a\in\R^3$
of the particle, $a,b = 1,2,3$. This point moves  along a trajectory in $\R^6$ defined by a Hamiltonian vector field $V_H^{}$ generated by a function $H(x,p)$ (Hamiltonian a.k.a. energy) with evolution parameter $t\in\R$ (time). For any choice of the Hamiltonian $H$, the space of initial data ($\equiv$ covariant phase space) determining the trajectory of motion coincides with the space $T^*\R^3=\R^6$.

 For a {\it relativistic} spinless particle, the non-relativistic phase space $\R^6$ is extended to space $T^*\R^{3,1}=\R^{3,1}\times\R^{3,1}$ with additional coordinates $x^0, p_0$, so that $(x^\mu )=(x^0, x^a)$ and $(p_\mu )=(p_0, p_a)$ are the coordinates and momenta of the relativistic particle. However, the space $T^*\R^{3,1}$ is not a covariant phase space in the definition of which the Hamiltonian function\footnote{In the relativistic case it is not particle's energy.} now participates. For example, a free relativistic particle is defined by a Hamiltonian function
\begin{equation}\label{1.1}
H_0=\frac{1}{2m}\,\eta^{\mu\nu}p_\mu p_\nu\quad\Rightarrow\quad V_{H_0}^{}=\frac{p^\mu}{m}\,\frac{\dpar}{\dpar x^\mu}=v^\mu\,\frac{\dpar}{\dpar x^\mu}\ ,
\end{equation}
generating the Hamiltonian vector field $V_{H_0}^{}$ and defining motion along straight lines in $T^*\R^{3,1}$,
\begin{equation}\label{1.2}
\dot x^\mu = V_{H_0}^{}x^\mu\ ,\quad\dot p_\mu = V_{H_0}^{}p_\mu\quad\Rightarrow\quad x^\mu (\tau )=x^\mu +v^\mu\tau\ ,\quad p_\mu (\tau )=p_\mu\ .
\end{equation}
Here $\dot x=\dd x/\dd\tau$ and $(\eta_{\mu\nu})=\diag (-1, 1, 1, 1)$ in \eqref{1.1} is the Minkowski metric. The evolution parameter $\tau$ in the general case does not coincide with the coordinate time $x^0$, but for free particles we have $x^0=\pm\tau$. On trajectories  \eqref{1.2} the function  \eqref{1.1} is constant, $H_0=-\sfrac12\,m$, and this equation (energy-momentum relation) defines a 7-dimensional hypersurface $H^3\times\R^{3,1}$ in the phase space $T^*\R^{3,1}$, where $H^3=H_+^3\cup H_-^3$ is a two-sheeted hyperboloid in momentum space. In this case, the initial data for the motion \eqref{1.2} are parametrized by the cotangent bundle $T^*H^3=H^3\times\R^{3,1}/\sGL(1,\R )$ defined by two equations:
\begin{equation}\label{1.3}
\eta^{\mu\nu} p_\mu p_\nu +m^2 =0\und p_\mu x^\mu =0\ .
\end{equation}
Here the group $\sGL(1,\R)\ni g=\exp (\tau V_{H_0}^{})$ generates the shifts along the trajectory \eqref{1.2}.

Note that the second equation  in \eqref{1.3} means that at the initial moment $\tau =0$ the particle trajectory is in the cotangent space $x^\mu = x^\mu_{\shortparallel}\in T^*_pH^3$ of the hyperboloid $H^3$, and $x^\mu_{\perp}=v^\mu\tau$ is a part of the solution \eqref{1.2} orthogonal to this hyperboloid, so that
\begin{equation}\label{1.4}
p_\mu (\tau)x^\mu(\tau)=p_\mu (x^\mu_{\shortparallel}+x^\mu_{\perp}) = p_\mu x^\mu + \frac{p_\mu p^\mu}{m}\tau = -m\tau\ .
\end{equation}
Geometrically this means that $x^\mu (\tau)=e^{\tau V_{H_0}^{}}x^\mu$ are orbits of the group $\sGL(1,\R)$ acting on the space $H^3\times\R^{3,1}$ and we have a principal bundle
\begin{equation}\label{1.5}
H^3\times\R^{3,1}\ \stackrel{\tiny\sGL(1,\R)}{\longrightarrow}\ T^*H^3
\end{equation}
with projection onto $T^*H^3$. Thus, the covariant phase space of a free relativistic particle is six-dimensional, as in the non-relativistic case, and a positive definite metric can be defined on it. The only difference is in the geometry: $T^*\R^3$ in the non-relativistic case and $T^*H^3$ in the relativistic case and besides, the Lorentz group maps the manifold $T^*H^3$ into itself.

Free spin-0 relativistic particles \eqref{1.1}-\eqref{1.5} in the Lagrangian approach can be described by a one-dimensional sigma model. To describe spin-$\sfrac12$ particles, fermionic (Grassmann odd) variables $\theta^\mu(\tau)$ should be added to this model as partners of the position variables $x^\mu(\tau)$ and it can be considered as a model of classical Dirac particles \cite{Ber, Brink}. The initial data $\theta^\mu = \theta^\mu(0)$ for the Grassmann variables satisfy the equation $\theta^\mu p_\mu =0$ (as does $x^\mu$), which defines an odd version $\Pi TH^3$ of the tangent bundle\footnote{In the presence of metric $\eta_{\mu\nu}$, one can ignore the difference between tangent and cotangent bundles.} $TH^3$. Here $\Pi$ is the operator which changes the Grassmann parity. Canonical quantization of these models of spin-0 and spin-$\sfrac12$ free relativistic particles leads to the Klein-Gordon and Dirac equations  \cite{Ber, Brink}, where the Dirac operator $\Gamma =\gamma^\mu\dpar_\mu$ can be written as an odd covariant Laplacian
\begin{equation}\label{1.6}
\Gamma =\gamma^\mu\frac{\dpar}{\dpar x^\mu}=\left(\eta^{\mu\nu}\frac{\dpar}{\dpar \theta^\nu}+\theta^\mu\right)\frac{\dpar}{\dpar x^\mu}
\end{equation}
on the space $\R^{3,1}\times\Pi\R^{3,1}$.

\medskip

\noindent {\bf Oscillating relativistic particles}. Quantum field theory is based on the free Klein-Gordon (KG) and Dirac equations, so a description of all facts regarding these equations and their solutions can be found in the textbooks. Less well known is the Klein-Gordon oscillator equation obtained by adding the Lorentz invariant function $V(x)=m^2\omega^2\eta_{\mu\nu}^{}x^\mu x^\nu$ to the KG equation (see e.g. \cite{Dirac}-\cite{Junker}), where $\omega$ is the angular frequency. This equation can be considered as a deformation of the KG equation since $\omega$ is a free parameter defining the external force and, when $\omega\to 0$, the KG oscillator equation is reduced to the free KG equation. The Dirac oscillator equation, which is a deformation of the free Dirac equation, was also introduced (see e.g. \cite{MV, Marr} and references therein).

The classical and quantum dynamics of the Klein-Gordon oscillator were considered in \cite{PopovJGP}. There the importance of considering the classical relativistic model of a particle and describing its covariant phase space was shown. The geometry of this space of initial data for motion of a relativistic particle depends on the Hamiltonian function $H(x,p)$, which distinguishes relativistic dynamics from non-relativistic ones. Constant value of $H$ fixes a 7-dimensional hypersurface $X_7\subset T^*\R^{3,1}=\R^{6,2}$ in phase space. This function $H$ also defines a Hamiltonian vector field $V_H$ generating a one-parameter group with elements $g=\exp (\tau V_H)$ acting on $X_7$. Here $\tau$ is a parameter on the orbits in $X_7$ along which the particle moves. The covariant phase space $X_6$ is obtained by quotienting $X_7$ by the action of this group. The non-trivial geometry of the space $X_6$ depends on $H$ and dictates the choice of an irreducible representation of the canonical commutation relations, i.e. in the relativistic case there is no analogue of the Stone-von Neumann theorem.

In \eqref{1.1}-\eqref{1.5} we discussed all the above steps for a free relativistic particle given by the Hamiltonian function \eqref{1.1} leading to the space $X_6=T^*H^3$. Interaction of this particle with an electromagnetic field $A=A_\mu\dd x^\mu$ is introduced by replacing $p_\mu$ by $P_\mu =p_\mu + eA_\mu$ in the function  $H_0$ from \eqref{1.1}. As a result we obtain a one-parameter family of 6-dimensional covariant phase spaces $X_6(e)$ whose geometry will differ from the cotangent bundle \eqref{1.3} even for very small $e\in \R$. This $X_6(e)$ is a non-integrable deformation of the free particle model $T^*H^3$, in contrast to the relativistic oscillator model, which is integrable for any value of the frequency parameter $\omega$. Its covariant phase space $Z_6$ and the transition from classical to quantum KG oscillator were described in detail in \cite{PopovJGP}. Below we recall these results and compare them with results for free particles. 

\medskip

\noindent {\bf Classical KG oscillator}. The classical relativistic oscillator is given by the Hamiltonian function
\begin{equation}\label{1.7}
H_\omega^{}=\frac{1}{2m}\bigl (\eta^{\mu\nu}p_\mu p_\nu + m^2\omega^2\eta_{\mu\nu}x^\mu x^\nu\bigr )\ \ \Rightarrow\ \ V_{H_\omega}^{}=\frac{p^\mu}{m}\frac{\dpar}{\dpar x^\mu}-m\omega^2x^\mu\frac{\dpar}{\dpar p^\mu}\ .
\end{equation}
The dynamics is given by the Hamiltonian vector field $V_{H_\omega}^{}$, which is the generator of the group $\sU(1)\ni g=\exp (\tau V_{H_\omega}^{})$ acting on the level surface
\begin{equation}\label{1.8}
H_\omega^{}=-\sfrac12\,m\ \Leftrightarrow\ \AdS_7:\ -2\omega^2\eta_{\mu\nb}z^\mu z^\nb =1\for 
z^\mu =\frac{1}{\sqrt{2}}\bigl (x^\mu - \frac{\im}{m\omega}p^\mu\bigr)\ .
\end{equation}
Covariant phase space is a homogeneous K\"ahler-Einstein manifold $Z_6$ obtained by quotienting $\AdS_7$ by the action of the dynamical group U(1),
\begin{equation}\label{1.9}
\AdS_7\ \stackrel{\sU(1)}{\longrightarrow}\ Z_6=\sSU(3,1)/\rS(\sU(3)\times\sU(1))\ ,
\end{equation}
which can be compared with the case of a free particle \eqref{1.5}. In the limit $\omega\to 0$, bundle \eqref{1.9} turns into bundle \eqref{1.5}.

Dynamical equations and their solutions have the form
\begin{equation}\label{1.10}
\dot z^\mu (\tau)=\im\omega z^\mu (\tau)\quad\Rightarrow\quad z^\mu (\tau)=e^{\tau V_{H_\omega}^{}}z^\mu =e^{\im\omega\tau}z^\mu\ ,
\end{equation}
where $z^\mu =z^\mu(0)$. From \eqref{1.10} it follows that
\begin{equation}\label{1.11}
x^\mu (\tau ) = x^\mu\cos\omega\tau + v^\mu\frac{\sin\omega\tau}{\omega}
\end{equation}
and therefore $x^0(\tau )\ne\pm\tau$ in contrast to the case of a free particle with $x^0=\pm\tau$. However, $\omega =2\pi/T$, where $T$ is the time for a single oscillation, and if we consider large $T$, for example comparable to the age of universe, the relativistic oscillator will be indistinguishable from free particles given in \eqref{1.2}. But even a very small $\omega$ changes the geometry of the covariant phase space of a particle, and instead of a disconnected manifold $T^*H^3$ having an infinite volume we obtain a simply connected manifold $Z_6$ having a finite volume, without limiting absolute values of coordinates and momenta. Note that the coordinate time $x^0$ does not coincide with the evolution parameter $\tau$ for any Hamiltonian function $H$ other than the function \eqref{1.1} defining a free particle. In particular, $x^0\ne\pm\tau$ for a particle interacting with an electromagnetic field.

On the manifold  $Z_6$ there is an almost complex structure $\CJ$ as well as a Riemannian metric. The K\"ahler-Einstein manifold $B_+^3:=(Z_6, \CJ)$ can be identified with the unit complex 3-ball in $\C^3$  with coordinates $y^a$,
\begin{equation}\label{1.12}
B_+^3=\left\{y^a:=\frac{z^a}{z^0}\ ,\  z^\mu\in\C^{3,1}\mid \delta_{a\bb}\,y^ay^\bb <1\right\}\ ,
\end{equation}
where the inequality on the right-hand side follows from the level set equation \eqref{1.8}. If we choose $p^0(\tau =0)>0$ in $B^3_+$, then in the conjugate manifold $B_-^3:=(Z_6, -\CJ)=\overline{B_+^3}$ with coordinates $y_-^a=y^\ab =\overline{y_+^a}$ we will have $p^0(\tau=0)<0$. Therefore, $B_+^3$ can be identified with the space of initial data for particles and $B_-^3$ with the space of initial data for antiparticles. In the limit $\omega\to 0$, the spaces $B^3_\pm$ are deformed in spaces $T^*H_\pm^3$. 

\medskip

\noindent {\bf Quantum KG oscillator}. Note that the coset space $Z_6$ from \eqref{1.9} and \eqref{1.12} is not a cotangent bundle over $x$-space, which explains various problems when considering the KG oscillator in the position representation. The space $(Z_6, \CJ )$ is K\"ahler, so these problems disappear when using the complex Segal-Bargman representation \cite{PopovJGP}. In this representation, the KG oscillator equation splits into two independent  equations for particles $\psi_+$ and antiparticles $\psi_-$ with general solutions of the form
\begin{equation}\label{1.13}
\psi_\pm^{}=e^{N/2}\left(\frac{1}{{\sqrt 2}\omega z_\pm^0}\right )^{N+2}f_\pm^{}(y^1_\pm , y^2_\pm , y^3_\pm)\for y_\pm^a = z_\pm^a/z_\pm^0 \ ,
\end{equation}
where $z_+^\mu =z^\mu\in\C^{3,1}=:\C^{3,1}_+, \ z_-^\mu =z^\mb=\overline{z_+^\mu}\in\C^{3,1}_-=\overline{\C^{3,1}},\ N=m/2\omega =1, 2, ...$ and $f_\pm^{}$ are arbitrary holomorphic functions of coordinates $y_\pm^{a}$ on $B_\pm^{3}$.

Holomorphic and antiholomorphic solutions $\psi_+$ and $\psi_-$ of the KG oscillator equation form weighted Bergman spaces which are Hilbert spaces of square-integrable holomorphic functions $f_\pm$ on $B_\pm^3$ defined as
\begin{equation}\label{1.14}
\CH_{\tt B}^\pm =L_h^2(B_\pm^3, \mu_{N+2}^{})=\Bigl\{\psi_\pm\ \mbox{from \eqref{1.13}}\mid\langle\psi_\pm , \psi_\pm\rangle = e^N\int_{B_\pm^3}^{}f_\pm^*f_\pm^{}\,\mu_{N+2}^{}\,\dd V_{\tt B}<\infty\Bigr\}\ ,
\end{equation}
where
\begin{equation}\label{1.15}
\mu_{N+2}^{}=\left ( \frac{1}{2\omega^2 z_\pm^0 z_\pm^{\bar 0}}\right)^{N+2}=\bigl (1-\delta_{a\bb}\,y_\pm^ay_\pm^\bb\bigr )^{N+2}
\end{equation}
is a weight function, $\dd V_{\tt B}=\im\varkappa^2\dd y^1\wedge\dd y^2\wedge\dd y^3\wedge\dd y^{\bar 1}\wedge\dd y^{\bar 2}\wedge\dd y^{\bar 3}$ and usually $\varkappa^2$ is chosen to be inversely proportional to the volume of the 3-balls $B_\pm^3$. Note that the functions $\psi_\pm^{}$ from \eqref{1.13} represent components of (3,0)-forms on $B_\pm^3$ with values in the complex line bundle $L_{N+2}^{}$ with the Hermitian metric given by the function \eqref{1.15}; this is their geometric meaning \cite{BH}.

The spaces \eqref{1.14} are  Hilbert spaces of unitary representation of the group SU(3,1) and its Lorentz subgroup SO(3,1). The bases in the weighted Bergman spaces \eqref{1.14} are given by functions\footnote{We did not care about normalization factors in eigenstates, integrals, etc., they are not important here.} 
\begin{equation}\label{1.16}
f_\pm (n_1, n_2, n_3) = (y_\pm^1)^{n_1}(y_\pm^2)^{n_2}(y_\pm^3)^{n_3}
\end{equation}
and their substitution into \eqref{1.13} yields eigenfunctions of the energy operator $\hat E$ with the energy eigenvalues\footnote{Throughout the paper, except for this formula, we use the natural units with $\hbar=c=1$.} \cite{PopovJGP}
\begin{equation}\label{1.17}
E(n_1, n_2, n_3) =mc^2\sqrt{1{+}\frac{2\hbar\omega}{mc^2}(n_1{+}n_2{+}n_3{+}\sfrac32)}\cong mc^2 + \hbar\omega(n_1{+}n_2{+}n_3{+}\sfrac32)\ \ \mbox{for}\ c^2\to\infty ,
\end{equation}
which are positive for all states of particles $\psi_+$ and antiparticles $\psi_-$. The creation and annihilation operators in this model are $y_\pm^a$ and $\dpar/\dpar y_\pm^a$. 

\medskip

\noindent {\bf Quantum relativistic particles.} Note that the quantum KG oscillator is described by the Hilbert space $\CH_{\tt B}^+$ of holomorphic functions on the covariant phase space $Z_6$ of the classical KG oscillator. This is exactly how the Hilbert space of non-relativistic quantum mechanics is associated with the covariant phase space $T^*\R^3$. The difference is that in the non-relativistic case the space of initial data $X_6=T^*\R^3$ does not depend on the Hamiltonian function, but in the relativistic case it does.

It is proposed to introduce a general rule: in the relativistic case, a quantum particle Hilbert space is always associated with (polarized) functions on the covariant phase space $X_6$ of the classical particle. If we accept this rule then the free Klein-Gordon field $\phi$ must be described by a quantum version of not only the first equation in  \eqref{1.3} but also a quantum version of the second equation in  \eqref{1.3}. Then in the momentum representation we obtain the equations
\begin{equation}\label{1.18}
(\eta^{\mu\nu}p_\mu p_\nu + m^2)\,\phi_\pm(p)=0\und \Bigl(p^\mu \frac{\dpar}{\dpar p_\mu }+ 2\Bigr)\,\phi_\pm(p)=0\ ,
\end{equation}
where $\phi_+$ is defined on the hyperboloid $H_+^3$ with $p^0>0$ and $\phi_-$ is defined on the hyperboloid $H_-^3$ with $p^0<0$. Recall that a free particle moves in accordance with equation \eqref{1.4} in the direction $x^\mu_\bot$ orthogonal to the covariant phase space \eqref{1.5}.

The general solutions of equations \eqref{1.18} are functions
\begin{equation}\label{1.19}
\phi_\pm (p)=\frac{m^2}{p_0^2}\vph_\pm(\xi_\pm^1, \xi_\pm^2, \xi_\pm^3)\ ,\quad \xi_\pm^a=\frac{p^a}{p^0_\pm}\with p^0_\pm=\pm|p^0|\ ,
\end{equation}
defined on the unit ball $H_\pm^3$ in $\R^3$,
\begin{equation}\label{1.20}
H_\pm^3=\Bigl\{\xi_\pm^a\in \R^3\mid w_3:=1-\delta_{ab}\,\xi_\pm^a\xi_\pm^{b}>0\Bigr\}\ ,
\end{equation}
which are the Beltrami-Klein models of real hyperbolic spaces $H_\pm^3$. Note that the second equation in \eqref{1.18} fixes the rescaling properties of functions $\phi_\pm(p)$ on $H_\pm^3$
 under momentum dilations $p_\mu\to\lambda p_\mu$, namely
\begin{equation}\label{1.21}
\phi_\pm^{}(\lambda p)=\lambda^{-2}\phi_\pm^{}(p)\ .
\end{equation}
The spaces $\CH^\pm_{\sf free}$ of such square-integrable functions with inner products
\begin{equation}\label{1.22}
\langle \phi_\pm^{} , \phi_\pm^{}\rangle= \int_{H_\pm^3}\phi_\pm^{*} \phi_\pm^{} \,\dd V_3^\pm = \int_{H_\pm^3}\vph_\pm^*\vph_\pm^{}\,w_3^2\,\dd V_3^\pm <\infty
\end{equation}
are real analogue of the weighted Bergman spaces \eqref{1.14}, $\dd V_3^\pm =\dd\xi_\pm^1\dd\xi_\pm^2\dd\xi_\pm^3$.

The basis (non-normalized) in the Hilbert space of functions \eqref{1.19}-\eqref{1.22} is given by functions
\begin{equation}\label{1.23}
\vph_\pm(n)=\bigl(\xi_\pm^1\bigr)^{n_1} \bigl(\xi_\pm^2\bigr)^{n_2} \bigl(\xi_\pm^3\bigr)^{n_3}\for 
(n)=(n_1, n_2, n_3)\ ,
\end{equation}
and their substitution into \eqref{1.19} yields eigenstates of the operator
\begin{equation}\label{1.24}
\hat D:=\frac\im 2 \bigl(\ph_0\xh^0+\xh^0\ph_0\bigr)=-\bigl(p^0\frac{\dpar}{\dpar p^0}+ \frac12\bigr)=p^a\frac{\dpar}{\dpar p^a}+ \frac32
\end{equation}
with the eigenvalues
\begin{equation}\label{1.25}
D(n)=n_1+n_2 +n_3 +\sfrac32\ .
\end{equation}
Thus, free particles $\phi_\pm (p)$ are parametrized by the momentum $p_\mu$ and quantum numbers $(n)$ given  by \eqref{1.23}-\eqref{1.25}.
The operators $\xi^a_\pm$ and ${\dpar}/{\dpar\xi^a_\pm}$ are operators of creation and annihilation acting  on the Hilbert spaces  $\CH^\pm_{\sf free}$ of free particles and antiparticles. The basis functions $\vph_\pm^{}(n)$ from \eqref{1.23} correspond to wave packets
\begin{equation}\label{1.26}
 \phi^\pm_{(n)} (x^0, x^a)= \int_{H_\pm^3}\vph_\pm^{}(n)\,w_3
e^{-\frac{\im m}{\sqrt{w_3}}(\pm x^0+\xi_a^\pm x^a)}\,\dd V_3^\pm\ ,
\end{equation}
characterized by quantum numbers \eqref{1.25}.

\medskip

\noindent {\bf Interaction picture}. Recall that the Hamiltonian \eqref{1.7} of the KG oscillator can be written as the Hamiltonian \eqref{1.1} of a free particle plus a perturbation with small $\omega$, and one can try to use the interaction representation to construct solution with $\omega\ne 0$ perturbatively. This will not work since
\begin{itemize}
\item covariant phase spaces $T^*H^3$ and $Z_6=\sU(3,1)/\sU(3)\times\sU(1)$ of free and oscillating relativistic particles are not diffeomorphic,
\item there is no homomorphism between quantum complex line bundles over $T^*H^3$ and $Z_6$,
\item there is no unitary map between creation  and annihilation operators in models of free and oscillating particles.
\end{itemize}
This distinguishes the relativistic case from the non-relativistic ones, where the covariant phase space is the same for perturbed and unperturbed cases. In fact, this is a general statement about the non-existence of the interaction picture in Lorentz covariant theories. Perhaps, this is what the Haag theorem of quantum field theory is connected with.

\medskip

\noindent {\bf Supersymmetric KG oscillator.} One of the main goals of this paper is to clarify the geometric meaning  of the Dirac equation as an odd covariant Laplacian on the space of initial data of spinning particles. We consider oscillating spinning particles, thus generalizing both the description of free spinning particles \cite{Ber, Brink} and the bosonic KG oscillator \cite{PopovJGP}. The model under consideration is a supersymmetric Klein-Gordon oscillator in Minkowski space with an equal number of bosonic and fermionic (Grassmann odd) coordinates in phase space. This model is defined by even and odd Hamiltonian functions $H_{\sf even}^{}$ and $H_{\sf odd}^{}$ generating vector fields $V_{H_{\sf even}}^{}$ and $V_{H_{\sf odd}}^{}$.
We will show that the covariant phase space of this oscillator is the odd tangent bundle $\Pi TZ_6$ of the manifold $Z_6$ from \eqref{1.9}, where the operator $\Pi$  changes the Grassmann parity of tangent spaces of $Z_6$. 

The quantum version of this model is given by the Dirac equation on phase space which is the odd covariant Laplacian on $\R^8\times\Pi\R^8=\Pi T(T^*\R^{3,1})$ acting on functions depending both on bosonic $(x^\mu , p_\mu)$ and fermionic $(\xi^\mu , \theta_\mu)$ coordinates. We show that the quantum Dirac and Klein-Gordon operators reduce to vector fields $V_{H_{\sf odd}}^{}$ and $V_{H_{\sf even}}^{}\sim V_{H_{\sf odd}}^{2}$ defining the classical supersymmetric KG oscillator.  We will show that solutions (components of spinor) of the quantum model 
are given by a set of holomorphic and antiholomorphic functions from the Bergman spaces \eqref{1.14} on $Z_6$ with weight functions $\mu_{N+k}^{}$ and $k=0, 1, 2, 3$.


\newpage

\noindent{\LARGE\bf Part I. Relativistic classical mechanics}

\section{Relativistic phase space of spinning particles}

\noindent {\bf Phase superspace}. Let us consider the phase space $T^*\R^{3,1}=\R^8$ of a relativistic spinless particle with coordinates $x^\mu\in\R^{3,1}$ and momenta $p_\mu\in\R^{3,1}$, $\mu , \nu =0,...,3$. We introduce coordinates
\begin{equation}\label{2.1}
x^{\mu +4}:=-w^2p^\mu = -w^2\eta^{\mu\nu}p_\nu\ ,
\end{equation}
where $w\in\R^+$ is a length parameter. We use the natural units with $\hbar =c=1$ so that $[w^2p^\mu]=[\mbox{length}]=[x^\mu]$. Let us consider the space $\R^8$ tangent to $T^*\R^{3,1}$ with basis $\{\dpar/\dpar x^M\}=\{\dpar/\dpar x^\mu , \dpar/\dpar x^{\mu +4}\}$ with $M=\{\mu , \mu +4\}=0,...,7$. Any vector $\xi$ in the tangent space has the form
\begin{equation}\label{2.2}
\xi =\xi^M\dpar_M=\xi^\mu\dpar_\mu + \xi^{\mu+4}\dpar_{\mu+4}\for \dpar_M:=\frac{\dpar}{\dpar x^M}\ .
\end{equation}
If we take as $\xi^M$ generators of the Grassmann algebra $\Lambda (\R^8)$, then the Grassmann-valued vectors \eqref{2.2} will be elements of the space $\Pi\R^8$, where the operator $\Pi$ inverts the Grassmann parity of the coordinates. Space
\begin{equation}\label{2.3}
\Pi T\R^8\cong\R^8\times\Pi\R^8\ni (x^M, \xi^M)
\end{equation}
is an odd tangent bundle of $\R^8$ with gr$\,x^M=0$ (parity) and gr$\,\xi^M=1$. It is the phase (super)space of relativistic  particles of spin $s=1/2$. For better consistency with \eqref{2.1} we introduce anticommuting coordinates
\begin{equation}\label{2.4}
\theta_\mu :=-\eta_{\mu\nu}\xi^{\nu +4}\quad\Rightarrow\quad\xi^{\mu +4}=-\eta^{\mu\nu}\theta_\nu\ .
\end{equation}
These variables $\xi^\mu$ and $\theta_\mu$ are dimensionless generators of the Grassmann algebra $\Lambda (\R^8)$, $[\xi^M]=L^0$.

\medskip

\noindent {\bf Symplectic structure.} The canonical symplectic structure on the space \eqref{2.3} is \cite{Kost}
\begin{equation}\label{2.5}
\begin{split}
\Omega &=\sfrac12\omega_{MN}^{\tt B}\dd x^M\wedge\dd x^N +\sfrac12 \omega_{MN}^{\tt F}\dd\xi^M\dd\xi^N\\[2pt]
&=\omega_{\mu\,\nu{+}4}^{\tt B}\dd x^\mu\wedge\dd x^{\nu +4}+\sfrac12 \omega_{\mu\nu}^{\tt F}\dd\xi^\mu\dd\xi^\nu +
\sfrac12 \omega_{\mu+4\,\nu+4}^{\tt F}\dd\xi^{\mu +4}\dd\xi^{\nu +4}\\[2pt]
&=\dd p_\mu\wedge\dd x^\mu + \sfrac12\eta_{\mu\nu}\dd\xi^\mu\dd\xi^\nu +\sfrac12\eta^{\mu\nu}\dd \theta_\mu\dd\theta_\nu\ ,
\end{split}
\end{equation}
where ``$\tt B$" and ``$\tt F$" mean ``bosonic" and ``fermionic". In defining differentials and derivatives for real coordinates \eqref{2.3} we follow Kostant \cite{Kost}, assuming
\begin{equation}\label{2.6}
\begin{split}
\dd\xi^M\dd\xi^N&=\dd\xi^N\dd\xi^M\ ,\quad\frac{\dpar}{\dpar\xi^M}\frac{\dpar}{\dpar\xi^N}=-\frac{\dpar}{\dpar\xi^N}\frac{\dpar}{\dpar\xi^M}\ ,
\\[2pt]
\dd x^M\dd\xi^N&=-\dd\xi^N\dd x^M\ ,\quad\frac{\dpar}{\dpar x^M}\frac{\dpar}{\dpar\xi^N}=\frac{\dpar}{\dpar\xi^N}\frac{\dpar}{\dpar x^M}\ ,
\\[2pt]
\xi^M\dd\xi^N&=-(\dd\xi^N)\xi^M\ ,\quad\xi^M\dd x^N=-(\dd x^N)\xi^M\ ,\quad 
x^M\dd \xi^N=(\dd \xi^N)x^M\ .
\end{split}
\end{equation}
From formulae \eqref{2.5} follow expressions for the components of the 2-form $\Omega$,
\begin{equation}\label{2.7}
\begin{split}
&\omega^{\tt B}_{\mu\,\nu+4}=\frac{1}{w^2}\eta_{\mu\nu}=-\omega^{\tt B}_{\nu+4\,\mu}\ ,
\\[2pt]
&\omega^{\tt F}_{\mu\nu}=\eta_{\mu\nu}=\omega^{\tt F}_{\nu\mu}\und\omega^{\tt F}_{\mu+4\,\nu+4}=\eta_{\mu\nu}=\omega^{\tt F}_{\nu+4\,\mu+4}\ .
\end{split}
\end{equation}
The Poisson structure on the graded symplectic space \eqref{2.3} has the form
\begin{equation}\label{2.8}
\{f,g\}=\omega_{\tt B}^{MN}\frac{\dpar f}{\dpar x^M}\frac{\dpar g}{\dpar x^N}+(-1)^{|f|}\omega_{\tt F}^{MN}\frac{\dpar f}{\dpar\xi^M}\frac{\dpar g}{\dpar\xi^N}\ ,
\end{equation}
where
\begin{equation}\label{2.9}
\omega_{\tt B}^{\mu\,\nu+4}=w^2\eta^{\mu\nu}=-\omega_{\tt B}^{\nu+4\,\mu}\ ,\quad 
\omega_{\tt F} ^{\mu\,\nu}=\eta^{\mu\nu}=\omega_{\tt F} ^{\nu\mu}\ ,\quad 
\omega_{\tt F} ^{\mu+4\,\nu+4}=\eta^{\mu\nu}=\omega_{\tt F} ^{\nu+4\,\mu+4}\ .
\end{equation}
Here $|f|=\mbox{gr}f=0$ for even functions $f$ and $|f|=\mbox{gr}f=1$ for odd functions $f$ \cite{Kost}.

\medskip

\noindent {\bf Complex coordinates.} We introduce on spaces $\R^8$ and $\Pi\R^8$ complex coordinates
\begin{equation}\label{2.10}
z^\mu =\frac{1}{\sqrt 2}(x^\mu + \im x^{\mu +4})\ ,\quad\eta^{\mu}=\frac{1}{\sqrt 2}(\xi^\mu + \im \xi^{\mu +4})
\end{equation}
and their complex conjugate
\begin{equation}\label{2.11}
z^{\bar \mu}=\frac{1}{\sqrt 2}(x^\mu - \im x^{\mu +4})\ ,\quad\eta^{\mb}=\frac{1}{\sqrt 2}(\xi^\mu - \im \xi^{\mu +4})
\end{equation}
with derivatives
\begin{equation}\label{2.12}
\begin{split}
\dpar_{z^\mu}^{}&=\frac{\dpar}{\dpar z^\mu}=
\frac{1}{\sqrt 2}\,\bigl (\dpar_\mu -\im\dpar_{\mu +4}\bigr)\ ,\quad
\dpar_{\eta^\mu}^{}=\frac{\dpar}{\dpar\eta^\mu}=\frac{1}{\sqrt 2}\,\bigl (\dpar_{\xi^\mu}^{}-\im\dpar_{\xi^{\mu +4}}^{}\bigr)\ ,\\[2pt]
\dpar_{z^\mb}^{}&=\frac{\dpar}{\dpar z^\mb}=
\frac{1}{\sqrt 2}\,\bigl (\dpar_\mu +\im\dpar_{\mu +4}\bigr)\ ,\quad
\dpar_{\eta^\mb}^{}=\frac{\dpar}{\dpar\eta^\mb}=\frac{1}{\sqrt 2}\,\bigl (\dpar_{\xi^\mu}^{}+\im\dpar_{\xi^{\mu +4}}^{}\bigr)\ .
\end{split}
\end{equation}
The symplectic 2-form \eqref{2.5} in these coordinates has the form
\begin{equation}\label{2.13}
\Omega=\frac{\im}{w^2}\eta_{\mu\nb}^{}\bigl (\dd z^\mu\wedge\dd z^\nb -\im w^2\dd\eta^\mu\dd \eta^\nb\bigr)\ .
\end{equation}
Note that we use the complex conjugation of the product of Grassmann numbers with a permutation: $(ab)^*=b^*a^*$.

\medskip

\noindent {\bf Tensor $\CJ$.} The derivatives in \eqref{2.12} form a basis of the tangent space $\CV$ to the space \eqref{2.3} and we can introduce an endomorphism $\tilde\CJ\in\,$End$(\CV)$ defined by formulae
\begin{equation}\label{2.14}
\begin{split}
&\tilde\CJ=\begin{pmatrix}\CJ&0\\0&\CJ\end{pmatrix}:\quad 
\begin{pmatrix}\R^8\\\Pi\R^8\end{pmatrix}\to\begin{pmatrix}\R^8\\\Pi\R^8\end{pmatrix}\ ,
\\[2pt]
&\CJ\Bigl(\frac{\dpar}{\dpar z^\mu}\Bigr){=}\im\frac{\dpar}{\dpar z^\mu} ,\ \
\CJ\Bigl(\frac{\dpar}{\dpar\eta^\mu}\Bigr){=}\im\frac{\dpar}{\dpar\eta^\mu},\ \
\CJ\Bigl(\frac{\dpar}{\dpar z^\mb}\Bigr){=}-\im\frac{\dpar}{\dpar z^\mb}\und
\CJ\Bigl(\frac{\dpar}{\dpar\eta^\mb}\Bigr){=}-\im\frac{\dpar}{\dpar \eta^\mb}\ .
\end{split}
\end{equation}
For the block diagonal matrix $\tilde\CJ$ we have an identity  $\tilde\CJ^2=-(\unit_8\oplus\unit_8)$ since we choose $\CJ^2=-\unit_8$. It is easy to see that
\begin{equation}\label{2.15}
\CJ\Bigl(\frac{\dpar}{\dpar x^M}\Bigr)=
\CJ_M^N\frac{\dpar}{\dpar x^N}\with
\CJ_\mu^{\nu +4}=\delta_\mu^\nu\und \CJ^\mu_{\nu +4}=-\delta^\mu_\nu\ ,
\end{equation}
and similarly
\begin{equation}\label{2.16}
\CJ\Bigl(\frac{\dpar}{\dpar\xi^M}\Bigr)=\CJ_M^N\frac{\dpar}{\dpar\xi^N}\ ,\quad
\CJ=\Bigl(\CJ_M^N\Bigr)\in\mbox{End}(\R^8)\ ,
\end{equation}
with the same tensor $\CJ$ of complex structure.

From \eqref{2.14}-\eqref{2.16} it follows that $(\R^8\times\Pi\R^8, \tilde\CJ)\cong\C^4\times\Pi\C^4$ is a complex space with complex coordinates \eqref{2.10} and $(\R^8\times\Pi\R^8, -\tilde\CJ)\cong\bar\C^4\times\Pi\bar\C^4$ is the complex conjugate space with coordinates \eqref{2.11}. With tensors $\CJ$ on spaces $\R^8$ and $\Pi\R^8$ one can associate vector fields
\begin{equation}\label{2.17}
\CJ_{\tt B}=\CJ_M^N x^M\frac{\dpar}{\dpar x^N}=\im z^\mu\dpar_{z^\mu}^{}-\im z^\mb\dpar_{z^\mb}^{}\und
\CJ_{\tt F} =\CJ_M^N \xi^M\dpar_{\xi^N}^{}=\im\eta^\mu\dpar_{\eta^\mu}^{}-\im \eta^\mb\dpar_{\eta^\mb}^{}
\end{equation}
defined on $\R^8$ and $\Pi\R^8$, respectively.

\medskip

\noindent {\bf Metric tensor}. Having a symplectic and complex structure on space $\R^8\times\Pi\R^8$, we can introduce a metric tensor with components given by the formulae
\begin{equation}\label{2.18}
\begin{split}
g_{MN}^{\tt B}&:=w^2\omega_{MK}^{\tt B}\CJ_N^K\quad\Rightarrow\quad g_{\mu\nu}^{\tt B}=\eta_{\mu\nu}^{}=g_{\mu+4\,\nu+4}^{\tt B}\ ,\\[2pt]
g_{MN}^{\tt F}&:=w^2\omega_{MK}^{\tt F}\CJ_N^K\quad\Rightarrow\quad g_{\mu\,\nu+4}^{\tt F}=-w^2\eta_{\mu\nu}^{}=-g_{\nu+4\,\mu}^{\tt F}\ .
\end{split}
\end{equation}
Using these formulae, we obtain the expression
\begin{equation}\label{2.19}
\begin{split}
g_{MN}^{\tt B}\dd x^M\dd x^N &+ g_{MN}^{\tt F}\dd \xi^M\dd \xi^N=\eta_{\mu\nu}^{}(\dd x^\mu\dd x^\nu+
\dd x^{\mu+4}\dd x^{\nu+4})\\[2pt]
&=\eta_{\mu\nu}^{}\dd x^\mu\dd x^\nu+w^4\eta^{\mu\nu}\dd p_\mu\dd p_\nu =:-\dd\tau^2\ ,
\end{split}
\end{equation}
for the infinitesimal interval on phase space, where $\tau$ parametrizes the trajectory of particle in phase space. This $\tau$ depends not only on relative velocity of particle but also on its acceleration. The terms with $\dd\xi^M$ dropped out of \eqref{2.19} due to the antisymetry of $g_{MN}^{\tt F}$. However,  $g_{MN}^{\tt F}\xi^M\xi^N$ is not equal to zero, which will be used later.

\section{Hamiltonian function of spinning particles}

\noindent {\bf Even Hamiltonian function $H_0$}. Using a metric on the phase space allows one to introduce Hamiltonian functions of particles using a scalar product. The simplest such function is a function of the form
\begin{equation}\label{3.1}
\begin{split}
H_0&=\frac{1}{2mw^4}\,\bigl(g_{MN}^{\tt B}x^Mx^N + g_{MN}^{\tt F}\xi^M\xi^N\bigr)=\frac{1}{2mw^4}\eta_{\mu\nu}\bigl(x^\mu x^\nu + x^{\mu +4}x^{\nu +4} -2w^2 \xi^{\mu}\xi^{\nu +4}\bigr)\\[2pt]
&=\frac{1}{mw^4}\eta_{\mu\nb}\bigl(z^\mu z^\nb  - \im w^2 \eta^{\mu}\eta^{\nb}\bigr)=\frac{1}{2m}\bigl(\eta^{\mu\nu}p_\mu p_\nu + m^2\omega^2\eta_{\mu\nu} x^{\mu}x^{\nu}+2m\omega \xi^{\mu}\theta_\mu\bigr)
\end{split}
\end{equation}
defining an even subsector of a supersymmetric relativistic oscillator\footnote{In the Introduction we used the notation $H_\omega^{}$, but now $H_0^{}$ denotes the Grassmann even Hamiltonian function, and the Hamiltonian function of a free particle will be denoted $H_0^{\rm free}$.}. Here
\begin{equation}\label{3.2}
\omega =\frac{1}{mw^2}=\frac{2\pi}{T}
\end{equation}
is the oscillator frequency, and $T$ is the period of a single oscillation. When $\xi^\mu =0=\theta_\mu$, the function 
\eqref{3.1} is the Hamiltonian function of the KG oscillator described in detail in \cite{PopovJGP}. When $T\to\infty$ ($\omega\to 0$) we obtain the function
\begin{equation}\label{3.3}
H_0^{\rm free} =\frac{1}{2m}\eta^{\mu\nu}p_\mu p_\nu
\end{equation}
defining a free Klein-Gordon particle.

Function \eqref{3.1} generates a Hamiltonian vector field
\begin{equation}\label{3.4}
\begin{split}
V_{H_0}^{}&=\{H_0,\cdot\}=\omega_{\tt B}^{MN}\frac{\dpar H_0}{\dpar x^M}\frac{\dpar }{\dpar x^N}+
\omega_{\tt F} ^{MN}\frac{\dpar H_0}{\dpar\xi^M}\frac{\dpar }{\dpar\xi^N}\\[3pt]
&=\omega\bigl (x^\mu\dpar_{x^{\mu{+}4}}^{}-x^{\mu +4}\dpar^{}_{x^{\mu}}+\xi^{\mu }\dpar_{\xi^{\mu{+}4}}^{}-\xi^{\mu{+}4}\dpar^{}_{\xi^{\mu}}\bigr)\\[3pt]
&=\im\omega\bigl(z^\mu\dpar_{z^\mu}^{}{-}z^\mb\dpar_{z^\mb}^{}{+}\eta^\mu\dpar_{\eta^\mu}^{}{-}\eta^\mb\dpar_{\eta^\mb}^{}\bigr)\\[3pt]
&=\frac{p^\mu}{m}\frac{\dpar }{\dpar x^\mu}-m\omega^2x^\mu\frac{\dpar }{\dpar p^\mu}+\omega\Bigl(\xi^\mu\frac{\dpar }{\dpar\theta^\mu}-\theta^\mu\frac{\dpar }{\dpar\xi^\mu}\Bigr).
\end{split}
\end{equation}
Comparing \eqref{2.17} and \eqref{3.4} we see that
\begin{equation}\label{3.5}
V_{H_0}^{}=\omega (\CJ_{\tt B}+\CJ_{\tt F} )=:\omega\tilde\CJ^v\ ,
\end{equation}
where $\CJ_{\tt B}$ is the generator of the group SO(2) of rotations in the plane $(x^\mu , x^{\mu +4})$, and $\CJ_{\tt F} $ is the generator of rotations in the plane $(\xi^\mu , \xi^{\mu +4})$. The function $H_0$ is constant on the orbits of this group in the phase space $\R^8\times\Pi\R^8$ since
\begin{equation}\label{3.6}
V_{H_0}^{}H_0=\{H_0, H_0\}=0
\end{equation}
by virtue of the definition of the vector field $V_{H_0}^{}$. Note that in the limit $\omega\to 0$ the vector field \eqref{3.4} transforms into the generator
\begin{equation}\label{3.7}
V_{H_0}^{\rm free}=\frac{p^\mu}{m}\frac{\dpar}{\dpar x^\mu}
\end{equation}
of the group GL(1, $\R$)=$\R^*$, as it should be for free particles.

\bigskip

\noindent {\bf Level set AdS$_7$}. Let us introduce the function
\begin{equation}\label{3.8}
\mu_{H_0}^{}=-2H_0=-2m\omega^2\eta_{\mu\nb}z^\mu z^\nb + 2\im\omega\eta_{\mu\nb}\eta^\mu \eta^\nb
\end{equation}
and consider its bosonic part $\mu^{\tt B}_{H_0}$ as a smooth map from space $\R^8\ni x^M$ to space $\R$ (momentum map \cite{MW}), the coordinate on which has dimension of $[p^\mu]=[m]=L^{-1}$, 
\begin{equation}\label{3.9}
\mu_{H_0}^{\tt B} : \ \R^8\to\R\with \mu_{H_0}^{\tt B} =-m\omega^2\eta_{\mu\nu}(x^\mu x^\nu + x^{\mu +4} x^{\nu +4})\ .
\end{equation}
The constant value $m>0$ of this function defines a hypersurface (a level set) in $T^*\R^{3,1}$,
\begin{equation}\label{3.10}
\mu_{H_0}^{\tt B} (x, p)=m\ :\quad\eta^{\mu\nu}p_\mu p_\nu+m^2\omega^2\eta_{\mu\nu}x^\mu x^\nu =-m^2\ ,
\end{equation}
which does not depend on the proper time $\tau$ due to \eqref{3.6} and coincides with the anti-de Sitter 
space $\AdS_7$ \cite{PopovJGP}.

Group U(1) with generator $\CJ_{\tt B}$ from \eqref{2.17} maps the manifold $\AdS_7$, embedded into $T^*\R^{3,1}$ via  \eqref{3.10}, into itself:
\begin{equation}\label{3.11}
\sU(1) \ni g=e^{\omega\tau\CJ_{\tt B}}: \quad (x^\mu , p^\mu ) \mapsto (x^\mu (\tau ), p^\mu (\tau ))\in \AdS_7\ .
\end{equation}
This Lie group action defines the dynamics -- motion of a particle along a circle $S^1$ in the fibres of the principal U(1)-bundle
\begin{equation}\label{3.12}
\AdS_7\ \stackrel{S^1}{\longrightarrow}\ Z_6\ ,\quad
Z_6=\AdS_7/\sU(1)\cong \sSU(3,1)/\rS(\sU(3)\times \sU(1))\ .
\end{equation}
The space of orbits $Z_6$ of this group is  the space of initial data of KG oscillator equations
\begin{equation}\label{3.13}
\dot z^\mu(\tau)=\omega\CJ_{\tt B}(z^\mu(\tau))=\im\omega z^\mu (\tau)\quad\Rightarrow\quad
z^\mu(\tau)=e^{\omega\tau\CJ_{\tt B}}z^\mu = e^{\im\omega\tau}z^\mu\ ,
\end{equation}
and this K\"ahler-Einstein space can be identified with the complex hyperbolic space $B_+^3=(Z_6, \CJ)$ for particles ($z^\mu$ with $p^0>0$ in $z^0$) or conjugate space $B_-^3=(Z_6, -\CJ)$  ($z^\mb$ with $p^0<0$ in $z^{\bar 0}$) for antiparticles \cite{PopovJGP}.

\bigskip

\noindent {\bf Level set $\Pi T\AdS_7$}. Given $(x,p)\in\R^8$, the differential of $\mu_{H_0}^{\tt B}$ at $(x,p)$ (the pushforward map) is a linear map
\begin{equation}\label{3.14}
\dd\mu_{H_0}^{\tt B} : \quad T_{(x, p)}\R^8\ \longrightarrow\ T_{\mu_{H_0}^{\tt B} (x, p)}\R
\end{equation}
from the tangent space of $\R^8$ at $(x,p)$ to the tangent space of $\R$ at $\mu_{H_0}^{\tt B} (x, p)=m$. Let us consider a vector
\begin{equation}\label{3.15}
\xi =\xi^M\frac{\dpar}{\dpar x^M}=\xi^\mu\frac{\dpar}{\dpar x^\mu} + \xi^{\mu+4}\frac{\dpar}{\dpar x^{\mu+4}}\in T_{(x,p)}^{}\R^8\ ,
\end{equation}
belonging to the space $T_{(x,p)}^{}\R^8$ at a point $(x^\mu , x^{\mu+4})$. The image of this vector under the map \eqref{3.14} is
\begin{equation}\label{3.16}
\dd\mu_{H_0}^{\tt B} (\xi)=-\frac{2}{mw^4}\eta_{\mu\nu}(\xi^\mu x^\nu + \xi^{\mu +4}x^{\nu +4})=\frac{2}{w^2}\,\zeta\ ,
\end{equation}
where $\zeta$ is the component of the vector in the tangent space $T_m\R$ to $\R$ at point $m$. Assuming $(\xi^M)=(\xi^\mu , \xi^{\mu +4})$ and $\zeta$ to be dimensionless Grassmann variables, we obtain the map
\begin{equation}\label{3.17}
\dd\mu_{H_0}^{\tt B} \ :\quad \Pi T^{}_{(x,p)}\R^8\ \longrightarrow\ \Pi T^{}_{m}\R
\end{equation}
of Grassmann odd tangent bundles. Next we consider $\xi^M$ as independent of $x^M$ and identify $\xi^M$ with the odd coordinates of the phase space \eqref{2.3}.

Let us now restrict $(x,p)\in\R^8$ to the manifold $\AdS_7$ given by equation \eqref{3.10} and in \eqref{3.16} let $\zeta =0$. Then equation \eqref{3.16} becomes the level set equation
\begin{equation}\label{3.18}
\mu_{H_1}^{}\equiv H_1:= \frac{1}{w^2}\,\eta_{\mu\nb}(\eta^\mu z^\nb + z^\mu\eta^\nb)=\theta^\mu p_\mu+m^2\omega^2x^\mu\xi_\mu =0\ ,
\end{equation}
defining the odd tangent bundle $\Pi T\AdS_7$ of the space $\AdS_7$. Accordingly, the space of initial data (covariant phase space) of oscillating particles is given by the odd tangent bundle $\Pi TZ_6$ for $Z_6=\AdS_7/\sU(1)$. Note that
equation \eqref{3.10} can be replaced by the constraint equation
\begin{equation}\label{3.19}
\mu_{H_0}^{}=\mu_{H_0}^{\tt B} +\mu_{H_0}^{\tt F}=\mu_{H_0}^{\tt B}  + 2\omega\,\theta_\mu\xi^\mu =m+2\omega\chi\ ,
\end{equation}
where $\mu_{H_0}^{\tt B} =m$ holds regardless of $\mu_{H_0}^{\tt F}=2\omega\chi$ since $\theta_\mu\xi^\mu $ and $\chi$ are imaginary even nilpotent variables. After quantization, constraint equations \eqref{3.18} and \eqref{3.19} lead to the equations of supersymmetric Klein-Gordon oscillator, which is the Dirac oscillator equation on the space $T^*\R^{3,1}$, and to the Klein-Gordon oscillator equations for the components of the spinor of the Dirac oscillator. 

In the limit $\omega\to 0$, equations \eqref{3.16}-\eqref{3.19} become equations for free spinning particles,
\begin{equation}\label{3.20}
\eta^{\mu\nu}p_\mu p_\nu +m^2=0\und \theta^\mu p_\mu + m\zeta =0\ ,
\end{equation}
defining an odd tangent bundle $\Pi TH^3$ (for $\zeta =0$) of a three-dimensional hyperboloid $p^2+m^2=0$ in the momentum space. Equations \eqref{3.20} were considered in \cite{Ber, Brink} within the framework of Lagrangian mechanics with constraints.

\medskip

\noindent {\bf Odd Hamiltonian function $H_1$.} In equation  \eqref{3.18} we defined a function $H_1$ on the phase space $\Pi T\R^8\cong\R^8\times\Pi\R^8$. This function generates a Hamiltonian vector field
\begin{equation}\label{3.21}
\begin{split}
V_{H_1}^{}&=\{H_1,\cdot\}=\omega_{\tt B}^{MN}\frac{\dpar H_1}{\dpar x^M}\frac{\dpar }{\dpar x^N}-
\omega_{\tt F} ^{MN}\frac{\dpar H_1}{\dpar\xi^M}\frac{\dpar }{\dpar\xi^N}\\[2pt]
&=\xi^\mu\dpar_{{\mu{+}4}}^{}-\xi^{\mu +4}\dpar^{}_{{\mu}}-\frac{1}{w^2}(x^{\mu }\dpar_{\xi^{\mu}}^{}
+x^{\mu{+}4}\dpar^{}_{\xi^{\mu+4}}\bigr)\\[2pt]
&=\im\eta^\mu\dpar_{z^\mu}^{}-\frac{1}{w^2}z^\mu\dpar_{\eta^\mu}^{}-\im\eta^\mb\dpar_{z^\mb}^{}{-}\frac{1}{w^2}z^\mb\dpar_{\eta^\mb}^{}\\[2pt]
&=\theta^\mu\frac{\dpar }{\dpar x^\mu}-p^\mu\frac{\dpar }{\dpar\theta^\mu}-m\omega\Bigl(x^\mu\frac{\dpar }{\dpar\xi^\mu}+\xi^\mu\frac{\dpar }{\dpar p^\mu}\Bigr)\ ,
\end{split}
\end{equation}
where we have used definition \eqref{2.8} with gr$\,H_1=|H_1|=1$. It is not difficult to verify that 
\begin{equation}\label{3.22}
V_{H_0}^{}H_1=\{H_0, H_1\}=0\ .
\end{equation}
This means that the function $H_1$ is constant on the orbits $S^1$ of the group U(1) generated by the vector field $V_{H_0}^{}=\omega\tilde\CJ^v$.

Recall that the vector field $\tilde\CJ^v=\CJ_{\tt B}+\CJ_{\tt F} $ from \eqref{2.17} and \eqref{3.5} has the form
\begin{equation}\label{3.23}
\tilde\CJ^v=\tilde\CJ^N_Mx^M\dpar_{x^M}^{}+\tilde\CJ^{N+8}_{M+8}\,\xi^M\dpar_{\xi^N}^{}\with
\tilde\CJ=\begin{pmatrix}\tilde\CJ^N_M&0\\0&\tilde\CJ^{N+8}_{M+8}\end{pmatrix}=
\begin{pmatrix}\CJ&0\\0&\CJ\end{pmatrix}\ .
\end{equation}
We introduce the ``square root" of matrix $\tilde\CJ$ as matrix
\begin{equation}\label{3.24}
\tilde\CJ_{1/2}^{}:=\begin{pmatrix}0&-w^{-1}\unit_8\\w\CJ&0\end{pmatrix}=
\begin{pmatrix}0&-w^{-1}\delta_M^N\\w\CJ_M^N&0\end{pmatrix}\quad\Rightarrow\quad
\tilde\CJ_{1/2}^{2}=-\tilde\CJ\ ,
\end{equation}
using which the vector field $V_{H_1}^{}$ can be written as
\begin{equation}\label{3.25}
V_{H_1}^{}=\frac{1}{w}(x^M\xi^M)\begin{pmatrix}0&-w^{-1}\delta_M^N\\w\CJ_M^N&0\end{pmatrix}
\begin{pmatrix}\dpar_{x^N}^{}\\\dpar_{\xi^N}^{} \end{pmatrix}=\CJ_M^N\xi^M\dpar_{x^N}^{}-\frac{1}{w^2}x^M\dpar_{\xi^M}^{}\ .
\end{equation}
It is easy to see that  identity \eqref{3.24} implies identity
\begin{equation}\label{3.26}
V_{H_1}^{2}=-mV_{H_0}^{}\ .
\end{equation}
In terms of the Poisson bracket \eqref{2.8}, identity \eqref{3.26} is equivalent to the identity
\begin{equation}\label{3.27}
\{H_1, H_1\}=-2mH_0\ .
\end{equation}
Note that ``$-m$" in identities \eqref{3.26} and \eqref{3.27} can be removed by redefining $H_1$ and $H_0$. Next we show that after quantization the function $H_0$ leads to the Klein-Gordon oscillator equation and $H_1$ to the Dirac oscillator equation in the Bargmann-Fock-Segal representation. Formulae \eqref{3.26} and \eqref{3.27} of classical mechanics are equivalent to the fact that the Dirac equation is the ``square root" of the Klein-Gordon equation.

\bigskip

\noindent {\bf Free particles}. Formulae \eqref{3.25}-\eqref{3.27} are preserved in the limit $w^2\to\infty$ ($\omega\to 0$) and we have
\begin{equation}\label{3.28}
\begin{split}
H_0^{\rm free}&=\frac{1}{2m}\eta^{\mu\nu}p_\mu p_\nu\ ,\quad 
V_{H_0}^{\rm free}=\frac{ p^\mu}{m}\frac{\dpar}{\dpar x^\mu}\ ,\\[2pt]
H_1^{\rm free}&=\theta^\mu p_\mu\ ,\quad
V_{H_1}^{\rm free}=\theta^\mu\frac{\dpar}{\dpar x^\mu}-p^\mu  \frac{\dpar}{\dpar \theta^\mu} ,\\[2pt]
\{H_1^{\rm free}, H_1^{\rm free}\}&=-2mH_0^{\rm free}\und 
\bigl(V_{H_1}^{\rm free}\bigr)^2=-p^\mu\frac{\dpar}{\dpar x^\mu}=-mV_{H_0}^{\rm free}\ .
\end{split}
\end{equation}
Next we will show that $H_0$ and $V_{H_0}^{}$ define a submanifold $Z_6\times\Pi\R^8$ in $\R^8\times\Pi\R^8$, and $H_1$ and $V_{H_1}^{}$ reduce it to an odd tangent bundle $\Pi TZ_6$ of dimension $(6|6)$. In the limit of free particles, this space of initial data for supersymmetric Klein-Gordon oscillators reduces to a graded space $T^*H^3\oplus\Pi TH^3$ of dimension $(6|3)$ with supersymmetry breaking.

\section{Equations of motion of spinning particles}

\noindent{\bf Dynamics}. The Hamiltonian functions $H_0$ and $H_1$ define a level surface in the initial phase space $\R^8\times\Pi\R^8$ and the particle must move in this submanifold. The equations of motion are defined in terms of the flows of Hamiltonian vector fields $V_{H_0}^{}$ and $V_{H_1}^{}$. In the case we are considering, the functions $H_0$ and $H_1$ define a submanifold $\Pi T\AdS_7$ in the superspace $\Pi T\R^8=\R^8\times\Pi\R^8$. This submanifold is given by equations
\begin{equation}\label{4.1}
H_0=-\sfrac12 m - \omega\chi\und H_1=-m\zeta\ ,
\end{equation}
where $\chi$ is an even nilpotent scalar and $\zeta$ is an odd Grassmann variable. The functions $H_0$ and $H_1$ define vector fields $V_{H_0}^{}$ and $V_{H_1}^{}$ that generate a superalgebra with commutators
\begin{equation}\label{4.2}
[V_{H_0}^{}, V_{H_0}^{}]_-=[V_{H_0}^{}, V_{H_1}^{}]_-=0
\und [V_{H_1}^{}, V_{H_1}^{}]_+=-2mV_{H_0}^{}\ .
\end{equation}
After quantization, $V_{H_0}^{}$ and $V_{H_1}^{}$ will be operators of the supersymmetric relativistic quantum mechanics defining the equations of the Klein-Gordon and Dirac oscillators, respectively.

The action of group U(1) with element $e^{\tau V_{H_0}}$ determines the dynamics of particles in proper time $\tau$. The flow equations given by $V_{H_0}$ are
\begin{equation}\label{4.3}
\begin{split}
\dot z^\mu =V_{H_0}^{}z^\mu =\im\omega z^\mu&\ \Rightarrow\ z^\mu (\tau )=e^{\tau V_{H_0}}z^\mu =e^{\im\omega\tau}z^\mu\ ,\\[2pt]
\dot \eta^\mu =V_{H_0}^{}\eta^\mu =\im\omega\eta^\mu&\ \Rightarrow\ \eta^\mu (\tau )=e^{\tau V_{H_0}}\eta^\mu =e^{\im\omega\tau}\eta^\mu\ ,
\end{split}
\end{equation}
where $z^\mu =z^\mu (0)$ and $\eta^\mu =\eta^\mu (0)$ are the initial data parametrizing the orbits \eqref{4.3} of group U(1). Here dot means the derivative $\dpar_\tau$. The group U(1) preserves constraint equations \eqref{4.1} since
\begin{equation}\label{4.4}
\dpar_\tau H_0 =V_{H_0}^{}H_0=0\und \dpar_\tau H_1 =V_{H_0}^{}H_1=0\ .
\end{equation}
The orbit space of this group is parametrized by the manifold $Z_6$.

It is easy to show that in the limit $\omega\to 0$, equations \eqref{4.3} are reduced to equations for free particles of the form
\begin{equation}\label{4.5}
\begin{split}
\dot x^\mu =V_{H_0}^{\rm free}x^\mu =\frac{p^\mu}{m}&\ \Rightarrow\ 
x^\mu (\tau)=x^\mu+\frac{p^\mu}{m}\tau\ ,\\[2pt]
\dot p^\mu =V_{H_0}^{\rm free}p^\mu =0&\ \Rightarrow\ 
p^\mu (\tau)=p^\mu=p^\mu(0)\ ,\\[2pt]
\dot\theta^\mu =V_{H_0}^{\rm free}\theta^\mu =0&\ \Rightarrow\ 
\theta^\mu (\tau)=\theta^\mu=\theta^\mu(0)\ .
\end{split}
\end{equation}
Note that these equations were previously derived in \cite{Brink} within the framework of the Lagrangian formalism.

\bigskip

\noindent{\bf Grassmann odd ``time".} Let us now consider the rotations given by the generator $V_{H_1}^{}$ of the two-parameter supergroup with generators \eqref{4.2}. One of the parameters of this group is the even ``time" $\tau$ with dynamics  \eqref{4.3}-\eqref{4.5}, and the second parameter is the Grassmann odd ``time" $\al$. The dynamics in even and odd times must preserve the space $\Pi T\AdS_7$, and for $\tau$ this is obvious from \eqref{4.4}.
For $\al$ we have
\begin{equation}\label{4.6}
\begin{split}
z^\mu\ &\mapsto\ \tilde z^\mu=z^\mu(\al)=e^{\frac{\al}{m}V_{H_1}}z^\mu=z^\mu+\frac{\im\al}{m}\eta^\mu\ \Leftrightarrow\ \frac{\dpar z^\mu}{\dpar\al}=\frac{\im}{m}\eta^\mu\ ,\\[2pt]
\eta^\mu\ &\mapsto\ \tilde \eta^\mu=\eta^\mu(\al)=e^{\frac{\al}{m}V_{H_1}}\eta^\mu=\eta^\mu+{\omega\al}z^\mu\ \Leftrightarrow\ \frac{\dpar\eta^\mu}{\dpar\al}=\omega z^\mu\ ,
\end{split}
\end{equation}
and it is easy to see that  $H_0$ is conserved, i.e.
\begin{equation}\label{4.7}
H_0\ \mapsto\ H_0(\al )= e^{\frac{\al}{m}V_{H_1}}H_0=H_0\ .
\end{equation}
Note that $\tau$ has the dimension of length, while the parameters $\al$, $\chi$ and $\zeta$ are chosen to be dimensionless.

It is easy to verify that the function $H_1$ is not invariant under transformations \eqref{4.6} but changes as follows:
\begin{equation}\label{4.8}
H_1\ \mapsto\ \tilde H_1=H_1(\al)=H_1-2\al H_0\ .
\end{equation}
Accordingly, for the level set \eqref{4.1} we obtain
\begin{equation}\label{4.9}
\tilde H_0=-\sfrac12(m+2\omega\chi )\und   \tilde H_1=-m\zeta+\al (m+2\omega\chi )\ ,
\end{equation}
which allows us to fix the dependence of Grassmann odd parameter $\zeta$ on $\al$. We choose
\begin{equation}\label{4.10}
\zeta =\al (1+\frac{2\omega}{m}\chi )\ ,
\end{equation}
that results in a level surface of the form
\begin{equation}\label{4.11}
\Pi T\AdS_7:\quad \tilde H_0=-\sfrac12(m+2\omega\chi )\und   \tilde H_1=0\ .
\end{equation}
By factorizing this surface under the action of the supergroup \eqref{4.2} with generators $V_{H_0}^{}$, $V_{H_1}^{}$, we obtain a covariant phase space $\Pi TZ_6$ diffeomorphic to the space $\Pi TB_+^3=\Pi T^{1,0}(Z_6, \CJ )$ or 
$\Pi TB_-^3=\Pi T^{1,0}(Z_6, -\CJ )$ depending on the choice of initial data with $p^0>0$ or $p^0<0$. Note that in \cite{Ber, Brink} supersymmetry transformations of the form \eqref{4.6} for a free particle were considered and the dependence on $\al$ was also fixed there.

\bigskip

\noindent{\bf Summary.} We have considered the phase space 
\begin{equation}\label{4.12}
\Pi T\R^8\cong\R^8\times\Pi\R^8\ni (x^M, \xi^M)=(x^\mu, p_\mu , \xi^\mu , \theta_\mu)
\end{equation}
of oscillating relativistic particles with spin $s=\sfrac12$ and introduced even and odd Hamiltonian functions $H_0$ and $H_1$. A constant value of the function $H_0$ defines a surface $\AdS_7\subset\R^8$, and the equality to zero of the function $H_1$ defines the reversed parity tangent\footnote{We usually speak of the tangent bundle because in the presence of a metric one can always identify tangent and cotangent bundles.} bundle $\Pi T\AdS_7$ on which the particle is located. Quotienting by the supergroup of rotations of bosonic and fermionic coordinates reduces the space $\Pi T\AdS_7$ to the space of initial data $\Pi TZ_6$ with $Z_6=\AdS_7/\sU(1)=\sU(3,1)/\sU(3)\times\sU(1)$, so that
\begin{equation}\label{4.13}
\Pi TZ_6\ \subset\ \Pi T\AdS_7\ \subset\ \Pi T\R^8\cong\R^8\times\Pi\R^8\ .
\end{equation}
On the space $Z_6$ one can introduce two conjugate complex structures $\pm\CJ$ and introduce the spaces $B_+^3=(Z_6, \CJ)$ and $B_-^3=(Z_6, -\CJ)$ of the bosonic parts of the initial data for particles and antiparticles. Adding fermionic initial data leads to supermanifolds $\Pi T^{1,0}Z_6\cong\Pi T B^3_+$ and $\Pi T^{0,1}Z_6\cong\Pi T B^3_-$ which are subbundles in the complexified odd tangent bundle $\Pi T^\C Z_6=\Pi T^{1,0}Z_6\oplus \Pi T^{0,1}Z_6$.

The variety $H^3_\C\equiv B_+^3$ (complex hyperbolic space) is a projectivization of the space $\C^{3,1}\cong\R^{6,2}=T^*\R^{3,1}$ and is covered by one patch, so that
\begin{equation}\label{4.14}
\C^{3,1}\ni z^\mu=(z^0, z^a) \to \left(1,\frac{z^a}{z^0}\right)=:(1, y^a)\in H^3_\C\ .
\end{equation}
From \eqref{3.10} it follows that $2z^0z^{\bar 0}\ge m^2w^4=\omega^{-2}$ and therefore the coordinates $y^a$ on complex hyperbolic space $H_\C^3$ must satisfy the condition
\begin{equation}\label{4.15}
\delta_{a\bar b}\,y^ay^{\bar b}<1\ ,
\end{equation}
i.e. they parametrize the open 3-ball in $\C^3$.

The dynamics is determined by evolution over $\tau$ with $\exp(\im\omega\tau)\in U(1)\subset\AdS_7$. We emphasize that the proper time $\tau$ is given by \eqref{2.19} and is a scalar with respect to the transformation group 
SU(3,1) of the phase space of KG oscillator. Coordinate time $x^0$ is a function of $\tau$ and coincides with $\pm\tau$ only in the limit of free particles. In this limit, the covariant phase spaces $\Pi T B_\pm^3$ of oscillating spinning particles and antiparticles are reduced to the covariant phase spaces of free spinning particles and antiparticles,
\begin{equation}\label{4.16}
\Pi T B_\pm^3\ \stackrel{\omega\to 0}{\longrightarrow}\ T^*H_\pm^3\oplus\Pi T H_\pm^3\ ,
\end{equation}
with dimensions (6$|$6) and (6$|$3), respectively. Due to the difference in the geometry of these supermanifolds, the quantized versions of the models of free and oscillating particles of spin $s=0$ and $s=\sfrac12$ differ significantly. We proceed to the description of quantum models in the next section.

\newpage

\noindent{\LARGE\bf Part II. Relativistic quantum mechanics}

\section{Quantum spinless particles}

\noindent{\bf Phase space.} The bosonic parts of supermanifolds \eqref{4.13} are manifolds
\begin{equation}\label{5.1}
H_\C^3=(Z_6, \CJ)\subset\AdS_7\subset T^*\R^{3,1}\ ,
\end{equation}
where $\CJ$ is a complex structure on $Z_6$ induced by the complex structure \eqref{2.15} on $\R^{6,2}\cong\C^{3,1}$. The complex unit ball $H_\C^3$ is described in  \eqref{4.14}, \eqref{4.15}. On the K\"ahler-Einstein coset space $H_\C^3\equiv B^3_+$ we will consider holomorphic and antiholomorphic functions. To unify formulae, it will be more convenient for us to consider antiholomorphic functions on $B_+^3$ as holomorphic functions on the conjugate manifold $B_-^3=(Z_6, -\CJ )=\overline{B_+^3}$ \cite{KobNom}. This manifold coincides with $B_+^3$ as a smooth manifold, $B_+^3\cong Z_6\cong B_-^3$.

Spaces \eqref{5.1} are associated with the classical relativistic oscillator. The transition to a quantum oscillator was described in detail in \cite{PopovJGP}. The supersymmetric relativistic oscillator is described by spaces \eqref{4.13}.
Its quantum version is described by the Dirac oscillator equation on phase space, the solutions of which will also satisfy the Klein-Gordon oscillator equations. Therefore, we will first describe the quantized Klein-Gordon oscillator, referring for details to \cite{PopovJGP}.

\medskip

\noindent{\bf  Bundles $L_\C^\pm$ and quantum charges}. By definition, the ``wavefunction" $\Psi_+$ of a particle is a section of a complex {\it vector} bundle $L_\C^+$ over $T^*\R^{3,1}$ of rank one, that is, its fibres are 
one-dimensional complex {\it vector} spaces $V^+\,{\cong}\,\C$ (see e.g.~\cite{Kost, Sni, Wood}). Therefore, $\Psi_+$ has the form $\Psi_+=\psi_+v_+$, where $v_+$ is a basis vector in $V^+$, and $\psi_+$ is a component of the vector-valued function $\Psi_+$ on $T^*\R^{3,1}$ with values in $V^+$. The fact that the ``wavefunction" is not a function but a vector in complex space is important for understanding quantum mechanics. If we consider only the bundle $L_\C^+$, then $v_+$ can be omitted and the component $\psi_+$ of the vector $\Psi_+$ will look like a function, although, unlike a scalar, it is transformed under automorphisms (gauge transformations) of the bundle $L_\C^+$. In general, if $\CE$ is a complex vector bundle, then there also exist the conjugate bundle $\bar\CE$ of $\CE$ which is obtained by having complex numbers acting through the complex conjugate of the numbers. These bundles are not isomorphic,  $\bar\CE$ has the first Chern class $c_1(\bar\CE)=-c_1(\CE)$ opposite to that of $\CE$. In our case the complex conjugate bundle is $L_\C^-:=\overline{L_\C^+}$ and both $L_\C^+$ and $L_\C^-$ can be obtained by the pull-back of conjugate line bundles over $(Z_6, \CJ)$. We must associate with $L_\C^+$ and $L_\C^-$ the charges $\qv =1$ and 
$\qv =-1$ called quantum charges in \cite{Popov}. These charges distinguish between particles and antiparticles. This description is obvious to everyone in the case of electromagnetic complex line bundles $E_\C^\pm$ with charges $\pm e$. However, electrically neutral particles also have quantum charge $\qv$, so bundles  $L_\C^\pm$ are not related to bundles $E_\C^\pm$.

The fibres of the bundle $L_\C^-$ are vector spaces $V^-\,{\cong}\,\bar{\C}$ with basis $v_-$, and sections of the bundle $L_{\C}^-$ are vector-valued functions $\Psi_-=\psi_-v_-$. The spaces $V^\pm$ are one-dimensional subspaces of the space $\C^2$ defined by the operator $J$,
\begin{equation}\label{5.2}
V^\pm:\quad Jv_\pm = \pm\im v_\pm\for J=\begin{pmatrix}0&-1\\ 1&0\end{pmatrix}\ 
\Rightarrow\ v_\pm =  \frac{1}{\sqrt 2}\,\begin{pmatrix}1\\ \mp\im\end{pmatrix}\ .
\end{equation}
Complex line bundles $L_\C^\pm$ are direct product spaces,
\begin{equation}\label{5.3}
L^\pm_\C = T^*\R^{3,1}\times V^\pm\ .
\end{equation}
Accordingly, sections $\Psi_\pm$ of bundles $L_\C^\pm$ are introduced as follows:
\begin{equation}\label{5.4}
\C^2\ni\Psi = \begin{pmatrix}\psi^1\\ \psi^2\end{pmatrix}=\Psi_+ + \Psi_- =\psi_+v_+ + \psi_-v_-\in V^+\oplus V^-\with 
\psi_\pm = \frac{1}{\sqrt 2}\,(\psi^1\pm\im\psi^2)\ .
\end{equation}
Here $\psi^1$ and $\psi^2$  are complex-valued, so $\psi_+$ and $\psi_-$ are not complex conjugate. They are complex conjugate for real-valued $\psi^1$ and $\psi^2$, and this in the case of particles with zero quantum charge (neutral like photons) described as sections of the bundle $L_\C^+\oplus L_\C^-$. Vector-valued functions $\Psi_+$ for particles and $\Psi_-$ for antiparticles can only be summed using a {\it direct sum}, as in \eqref{5.4}.

The matrix $J$ introduced in \eqref{5.2} is the generator of the structure group U(1)$_{\sf v}\cong \sSO (2)_{\sf v}$ of complex line bundles $L_\C^\pm$. Associated with this group is the above-mentioned quantum charge $\qv=\pm 1$, which is an eigenvalue of the operator $Q_{\sf v}$,
\begin{equation}\label{5.5}
Q_{\sf v}:= -\im J:\quad Q_{\sf v}\Psi_\pm = \qv\Psi_\pm =\pm\Psi_\pm \ ,
\end{equation}
for $\Psi_\pm\in\Gamma (T^*\R^{3,1}, L_\C^\pm)$. It is convenient to combine the bundles $L_\C^+$ and $L_\C^-$
into a vector bndle $L_{\C^2}^{}=L_\C^+\oplus L_\C^-$ of rank 2. The connection one-form $\Av$ on this bundle is given by the one-form $\theta_{\R^8}$ of symplectic potential on $T^*\R^{3,1}$,
\begin{equation}\label{5.6}
\begin{split}
\Av =\theta_{\R^8}^{}J=\frac12\,(p_\mu\dd x^\mu - x^\mu\dd p_\mu )J=\frac{1}{2w^2}\,\eta_{\mu\nu}(x^\mu\dd x^{\nu +4}- x^{\mu +4}\dd x^\nu )J\\[2pt]
=\frac12\,m\omega\eta_{\mu\nb}(z^\nb\dd z^\mu - z^\mu\dd z^\nb)J\quad \Rightarrow\quad \Fv=\dd\Av =\omega^{\tt B}J\ ,
\end{split}
\end{equation}
where the bosonic part $\omega^{\tt B}$ of the symplectic 2-form $\Omega$ on $T^*\R^{3,1}$ is written out in \eqref{2.5} and \eqref{2.13}. The abbreviation ``$\sf v$" and ``$\sf vac$" here mean ``$\sf vacuum$" since $\theta_{\R^8}^{}$ and $\omega^{\tt B}$ have no sources and define the canonical symplectic structure on $\R^8$.

\bigskip

\noindent{\bf Polarizations and CCR.} To move to relativistic quantum mechanics, it is necessary to introduce spaces of sections $\CF_{\tt B}^\pm$ of bundles $L_\C^\pm$ on which an irreducible representation of the canonical commutation relations (CCR) can be defined. We will use the Bargmann-Fock-Segal representation \cite{Segal, Barg, Hall}, in which we need to define holomorphic structures in Hermitian bundles $L_\C^\pm$. It is convenient for us to introduce notation $z_+^\mu :=z^\mu$ and $z_-^\mu :=\overline{z^\mu_+}=z^\mb$ for holomorphic coordinates on the spaces 
$\C^{3,1}_+=\C^{3,1}\cong T^*\R^{3,1}$ and $\C^{3,1}_-=\overline{\C^{3,1}}$. Using $z_\pm^\mu$ we define the Dolbeault operators
\begin{equation}\label{5.7}
\bar\dpar^{}_{L_\C^\pm} = \dd z_\pm^{\bar\mu}\, \left (\frac{\dpar}{\dpar z_\pm^{\bar\mu}}+\frac{1}{2w^2}\eta_{\bar\mu\nu}z^\nu_\pm\right )
\end{equation}
and impose on $\Psi_\pm$ the polarization conditions
\begin{equation}\label{5.8}
\bar\dpar^{}_{L_\C^\pm}\Psi_\pm =0\ .
\end{equation}
Solutions of these constraint equations are 
\begin{equation}\label{5.9}
\Psi_\pm =\psi_\pm (z^\mu_\pm, \tau )v_\pm^c\with v_\pm^c=\psi_0^{\tt B}(z_\pm , \bar z_\pm ) v_\pm\ ,
\end{equation}
where 
\begin{equation}\label{5.10}
\psi_0^{\tt B}=\exp\Bigl(-\frac{1}{2w^2}\eta_{\mu\bar\nu}z^\mu_\pm z_\pm^{\bar \nu}\Bigr )
=\exp\Bigl(-\frac{1}{2w^2}\eta_{\mu\bar\nu}z^\mu z^{\bar \nu}\Bigr )\ .
\end{equation}
Note that $\psi_+$ are holomorphic and $\psi_-$ are antiholomorphic functions of $z^\mu$. These conditions replace the conditions of positive and negative frequency functions for particles and antiparticles in position representation.

\bigskip

\noindent{\bf Covariant derivatives.} A connection $\Av$ in the bundle $L_{\C^2}^{}$ defines complex conjugate connections $A^\pm_{\sf vac}$ in bundles $L_{\C}^{\pm}$ since $J=\pm\im$ on $L_{\C}^{\pm}$. However, in coordinates $z_\pm^\mu$ these connections have the same form and define covariant derivatives of the form
\begin{equation}\label{5.11}
\nabla_{z^\mu_\pm}=\dpar_{z^\mu_\pm} +  A_{z^\mu_\pm}^\pm= \dpar_{z^\mu_\pm} +  \frac{1}{2w^2}\eta_{\mu\bar\nu}z_\pm^{\bar \nu}\ ,
\ \ \nabla_{z_\pm^{\bar \mu}}=\dpar_{z_\pm^{\bar \mu}} + A_{z_\pm^{\bar \mu}}^\pm=\dpar_{z_\pm^{\bar \mu}}-\frac{1}{2w^2}\eta_{\bar\mu\nu}z_\pm^{\nu}\ .
\end{equation}
It is easy to verify that on polarized sections \eqref{5.9} of bundles $L_{\C}^{\pm}$ these operators of covariant derivatives have the form
\begin{equation}\label{5.12}
\nabla_{z_\pm^\mu}^{}\Psi_\pm^{} = \Bigl(\dpar_{z_\pm^\mu}^{}\psi_{\pm}^{}\Bigr)v_\pm^c\und \nabla_{z_\pm^{\bar \mu}}^{}\Psi_\pm^{} =-\Bigl(\frac{1}{w^2}\eta_{\bar\mu\nu}^{} z_\pm^\nu\psi_\pm^{}\Bigr)v_\pm^c\ ,
\end{equation}
where $v_\pm^c$ are given in \eqref{5.9}. From \eqref{5.12} it follows that the annihilation and creation operators for sections $\Psi_\pm\in\Gamma(L_\C^\pm)$ have the form
\begin{equation}\label{5.13}
a_{\mu\pm}^{}=w\nabla_{z_\pm^\mu}^{}\und a_{\bar\mu\pm}^\+=-w\nabla_{z_\pm^{\bar \mu}}^{}\with [a_{\mu\pm}^{}, a^\+_{\bar\nu\pm}]=\eta_{\mu\bar\nu}^{}\ .
\end{equation}
Thus, we have introduced Fock spaces $\CF_{\tt B}^\pm$ of functions $\psi_\pm $ from \eqref{5.9} holomorphic on spaces 
$\C_\pm^{3,1}$ and taking values in $V^\pm$.

\medskip

\noindent{\bf AdS$_7$ energy shell.} On the spaces $\CF_{\tt B}^\pm$, irreducible representations of CCR \eqref{5.13} are realized, as well as a representation of the group SU(3,1). These representations are not unitary due to the presence of $\eta_{0\bar 0}^{}=-1$ in the function $\psi_0^{\tt B}$ from \eqref{5.10} and in the commutators \eqref{5.13}. However, the coordinates and momenta of the oscillating particles we are considering must satisfy the relativistic constraint equation
\begin{equation}\label{5.14}
-\frac{2}{w^4}\eta_{\mu\nb}^{}z^\mu z^\nb =m^2\ .
\end{equation}
In fact, any Lorentz-invariant Hamiltonian function $H$ defines a 7-dimensional hypersurface $X_7$ in $T^*\R^{3,1}$ on which the  particle must be located. On the surface \eqref{5.14} the function \eqref{5.10} is constant, $\psi_0^{\tt B}=\exp(m/4\omega )$. The motion of a particle in $\AdS_7$-space \eqref{5.14} is parametrized by the K\"ahler-Einstein manifold $Z_6$ and we will show that solutions of the Klein-Gordon oscillator equation are given by functions on this space that does not contain $\eta_{0\bar 0}$ from \eqref{5.11}-\eqref{5.13}.

\medskip

\noindent{\bf Evolution equations.} So, we have holomorphic sections \eqref{5.9} of  the bundles $L_\C^\pm$ and covariant derivatives \eqref{5.11} acting on $\Psi_\pm$ as shown in \eqref{5.12}.  Quantization is a transition from the phase space $T^*\R^{3,1}$ with Hamiltonian function $H_0$ and a level set $\mu_{H_0}^{\tt B}=m$ to the bundle $L_{\C^2}^{}$ and an operator $\hat\mu_{H_0}^{\tt B}$ acting on sections of this bundle. We introduce an operator version of the momentum map \eqref{3.9} as
\begin{equation}\label{5.15}
\hat\mu_{H_0}^{\tt B} = \frac{1}{m}\,\Delta_2=\frac{1}{m}\,\eta^{\mu\nb}(\nabla_\mu^{\tt B}\nabla_\nb^{\tt B}+\nabla_\nb^{\tt B}\nabla_\mu^{\tt B})
\ ,
\end{equation}
where $\Delta_2$ is the covariant Laplacian acting on sections $\Psi =\Psi_++\Psi_-$ of the bundle $L_{\C^2}^{}=
L_\C^+\oplus L_\C^-$. The covariant derivatives in \eqref{5.15} are
\begin{equation}\label{5.16}
\nabla_\mu^{\tt B}=\dpar_{z^\mu}^{} + \frac{1}{2w^2}\eta_{\mu\nb}z^\nb Q_{\sf v}\und 
\nabla_\mb^{\tt B}=\dpar_{z^\mb}^{} - \frac{1}{2w^2}\eta_{\mb\nu}z^\nu Q_{\sf v}\ ,
\end{equation}
where $Q_{\sf v}=-\im J$ is the matrix of quantum charge, $Q_{\sf v}\Psi_\pm =\pm\Psi_\pm$. When restricted to $L_\C^+$ and $L_\C^-$, formulae \eqref{5.16} take the form \eqref{5.11} after replacing $Q_{\sf v}$ with $\pm 1$ on $L_\C^\pm$.

Bundles $L_\C^+$ and $L_\C^-$ are complex conjugate, so the evolution in $\tau$ of $\Psi_+$ and $\Psi_-$ have to be conjugate. We will introduce evolution of $\Psi =\Psi_++\Psi_-$ by the equation
\begin{equation}\label{5.17}
J\dpar_\tau\Psi = \sfrac1m\,\Delta_2\Psi\quad\Rightarrow\quad\pm\im\dpar_\tau\Psi_\pm= \sfrac1m\,\Delta_2\Psi_\pm\ . 
\end{equation}
It is not difficult to show that two independent continuity equations follow from these equations:
\begin{equation}\label{5.18}
\dpar_\tau^{}\rho_\pm^{}+\nabla_{z^\mu}^{}j_\pm^\mu +\nabla_{z^\mb}^{}j_\pm^\mb =0\ ,
\end{equation}
where $\rho_\pm^{}=\pm\Psi_\pm^\+\Psi_\pm$ and
\begin{equation}\label{5.19}
 j_\pm^\mu:=
\frac{\im}{m}\,\eta^{\mu\nb}\Bigl(\Psi_\pm^\+\nabla_{z^{\nb}}\Psi_\pm - (\nabla_{z^{\nb}}\Psi_\pm^\+)\Psi_\pm\Bigr)\ ,
\quad
j^\mb_\pm :=\frac{\im}{m}\,\eta^{\mb\nu}\Bigl(\Psi_\pm^\+\nabla_{z^\nu}^{}\Psi_\pm - 
(\nabla_{z^\nu}^{}\Psi_\pm^\+)\Psi_\pm\Bigr)\ .
\end{equation}
The quantities $\rho_\pm^{}$ are the densities of quantum charges, and $\Psi_\pm^\+\Psi_\pm$ can be related to probability densities. 

We will choose the dependence of $\Psi$ on $\tau$ in the form
\begin{equation}\label{5.20}
\Psi =e^{-\im E_0\tau}\Psi_+   + e^{\im E_0\tau}\Psi_- \with\Psi_\pm =\psi_\pm(z_\pm )\,v_\pm^c\ ,
\end{equation}
where $E_0$ is an arbitrary energy parameter. Substituting \eqref{5.20} into \eqref{5.17}, we obtain equations for relativistic oscillators
\begin{equation}\label{5.21}
(\Delta_2-m E_0 )\,\Psi_\pm =0\ ,
\end{equation}
where $\Psi_\pm$ does not depend on $\tau$. For the KG oscillator, $E_0$ is not arbitrary, but is fixed at $E_0=m$, and we have the equation of the Klein-Gordon oscillator,
\begin{equation}\label{5.22}
(\Delta_2-m^2)\,\Psi_\pm =0\ .
\end{equation}
It is not difficult to show that after substituting $\Psi_\pm$ from \eqref{5.9}-\eqref{5.12} into \eqref{5.22}, the KG oscillator equations are reduced to the equations
\begin{equation}\label{5.23}
\Bigl(z_\pm^\mu\frac{\dpar}{\dpar z_\pm^\mu}+N+2\Bigl)\,\psi_\pm(z_\pm)=0\ ,
\end{equation}
where
\begin{equation}\label{5.24}
N=\frac{m^2w^2}{2}=\frac{m}{2\omega}>0\ .
\end{equation}
We will consider $N$ as an interger fixed by the choice of the parameter $\omega$.

\medskip

\noindent{\bf Solution space $\CH_{\tt B}=\CH_{\tt B}^+\oplus\CH_{\tt B}^-$.} We will discuss only solutions $\Psi_+\in\CH_{\tt B}^+$ of equation \eqref{5.22} since for $\Psi_-\in\CH_{\tt B}^-$ everything is similar. The general solution of equation \eqref{5.23}  for $\psi_+$ is
\begin{equation}\label{5.25}
\psi_+=\biggl(\frac{1}{\sqrt 2\omega z^0}\biggr)^{N+2}f_+(y^1 , y^2 , y^3 )\ ,
\end{equation}
where $f_+$ is an arbitrary holomorphic function of the coordinates $y^a=z^a/z^0$ on the unit complex 3-ball $H_\C^3$ in $\C^3$ given by equation \eqref{4.15}.

Recall that the coordinates and momenta of the oscillator are arbitrary, they are restricted only by equation \eqref{5.14}  (energy shell). From \eqref{5.14} it follows that 
\begin{equation}\label{5.26}
\biggl(\frac{1}{ 2\omega^2 z^0 z^{\bar 0}}\biggr)^{N+2}=
(1-\delta_{a\bb}\,y^a y^\bb)^{N+2}=:\mu_{N+2}^{}\ ,
\end{equation}
and therefore
\begin{equation}\label{5.27}
\Psi_+^\+\Psi_+^{}=(\psi_+^{}\psi_0^{\tt B}\, v_+^{})^\+(\psi_+^{}\psi_0^{\tt B}\, v_+^{})=e^N f_+^*f_+^{}\mu_{N+2}^{}\ ,
\end{equation}
where $N$ is defined in \eqref{5.24}. For two different solutions $\Psi$ and $\hat\Psi$ of equation \eqref{5.22} the inner product is defined as
\begin{equation}\label{5.28}
\langle\Psi_+^{} , \hat\Psi_+^{}\rangle := \int_{B_+^3}\Psi_+^\+\hat\Psi_+^{}\dd V_ {\tt B}= e^N\int_{B_+^3}f_+^*\hat f_+^{}\,\mu_{N+2}^{}\,\dd V_{\tt B}\ ,
\end{equation}
where $\dd V_{\tt B}=\im\vk^2\dd y^1\wedge\dd y^2\wedge\dd y^3\wedge\dd y^{\bar 1}\wedge\dd y^{\bar 2}\wedge\dd y^{\bar 3}$ and usually $\vk^2$ is chosen to be inversely proportional to the volume of the 3-balls $B_\pm^3$. From \eqref{5.25}-\eqref{5.28} we conclude that the space $\CH_{\tt B}^+$ of all holomorphic  solutions of the Klein-Gordon oscillator equation \eqref{5.22} is the weighted Bergman space
\begin{equation}\label{5.29}
\CH^+_{\tt B} = L_h^2(B_+^3, \mu_{N+2}^{})=\{\Psi_+^{}\in\CF^+_{\tt B}\ \mbox{with}\ \eqref{5.25}{-}\eqref{5.28}\mid
\langle\Psi_+^{} , \Psi_+^{}\rangle<\infty\}
\end{equation}
with the measure $\mu_{N+2}^{}$. Antiholomorphic solutions of the KG oscillator equation belong to the Bergman space $\CH^-_{\tt B} = L_h^2(B_-^3, \mu_{N+2}^{})$ and all solutions are given by the {\it direct sum} of these spaces $\CH_{\tt B}=\CH_{\tt B}^+\oplus\CH_{\tt B}^-$. For more details on weighted Bergman spaces see e.g. \cite{BH, ZZ} and references therein.

\section{Quantum spinning particles}

\noindent{\bf Sheaves $\CL_\C^\pm$.} In Section 4 we described the phase space $\R^8\times\Pi\R^8$ of spinning particles and its reduction to the covariant phase space $\Pi TZ_6$. To describe quantum particles, we must first define bundles $L_\C^\pm$ over $\R^8\subset\R^8\times\Pi\R^8$ with spaces of sections $\Gamma (\R^8, L_\C^\pm)\cong C^\infty (\R^8, V^\pm)$, which are spaces of functions on $\R^8$ with values in vector spaces $V^\pm$. Secondly, we need to extend $L_\C^\pm$ to $\Pi\R^8$. Line bundles over $\Pi\R^8$ are not defined since $\Pi\R^8$ does not exist as a set of points. However, any locally trivial bundle is equivalently described by the sheaf of its sections. Therefore, instead of complex line bundle over $\Pi\R^8$, one should consider the complexified space of functions on $\Pi\R^8$ that form a Grassmann algebra $\Lambda (\C^8)=\Lambda (\R^8)\otimes\C$ of polynomial in eight Grassmann variables. Thus, instead of  complex line bundles over $\R^8\times\Pi\R^8$, we will consider the space of functions on $\R^8\times\Pi\R^8$ with values in $\C^2=V^+\oplus V^-$ which can be identified with the space
\begin{equation}\label{6.1}
\CL_{\C^2}^{}=\Gamma (\R^8, L_{\C^2}^{})\otimes\Lambda (\R^8)=\CL_{\C}^{+}\oplus\CL_{\C}^{-}\with
\CL_{\C}^{\pm}=\Gamma (\R^8, L_{\C}^{\pm})\otimes\Lambda (\R^8)
\end{equation}
of smooth functions on $\R^8$ with values in the vector space $V^+\otimes\Lambda (\R^8)\oplus V^-\otimes\Lambda (\R^8)$. Following Kostant \cite{Kost}, we will call the spaces $\CL_{\C}^{\pm}$ line bundle sheaves.

\bigskip

\noindent{\bf Covariant derivatives.} To differentiate functions from spaces $\CL_{\C}^{\pm}$, the vector fields $\dpar_{x^M}^{}$ and $\dpar_{\xi^M}^{}$ must be replaced by covariant derivatives
\begin{equation}\label{6.2}
\nabla_{x^M}^{\pm}=\dpar_{x^M}^{}+A_{x^M}^{\pm}\und 
\nabla_{\xi^M}^{\pm}=\dpar_{\xi^M}^{}+A_{\xi^M}^{\pm}\ ,
\end{equation}
where
\begin{equation}\label{6.3}
A_{\sf vac}^\pm = A_{x^M}^{\pm}\dd x^M + A_{\xi^M}^{\pm}\dd \xi^M
\end{equation}
are connections on sheaves $\CL_{\C}^{\pm}$ such that their curvatures $F_{\sf vac}^\pm$ are proportional to the symplectic 2-form $\Omega$ from \eqref{2.5} \cite{Kost},
\begin{equation}\label{6.4}
F_{\sf vac}^\pm=\dd A_{\sf vac}^\pm =\pm\im\Omega\ .
\end{equation}
The bosonic part of the connection in \eqref{6.3} was described in detail in Section 5. The fermionic part of the connection \eqref{6.3} in complex Grassmann variables has the form
\begin{equation}\label{6.5}
A_{\eta^\mu}^\pm =\pm\sfrac\im 2\,\eta_{\mu\nb}^{}\eta^\nb\und
A_{\eta^\mb}^\pm =\pm\sfrac\im 2\,\eta_{\mb\nu}^{}\eta^\nu
\end{equation}
and therefore
\begin{equation}\label{6.6}
\nabla_{\eta^\mu}^\pm =\dpar_{\eta^\mu}+ A_{\eta^\mu}^\pm =\dpar_{\eta^\mu}\pm\sfrac\im 2\,\eta_{\mu\nb}^{}\eta^\nb\ ,\quad
\nabla_{\eta^\mb}^\pm =\dpar_{\eta^\mb}+  A_{\eta^\mb}^\pm=
\dpar_{\eta^\mb}\pm\sfrac\im 2\,\eta_{\mb\nu}^{}\eta^\nu\ .
\end{equation}
It is easy to verify that non-vanishing anticommutators of operators \eqref{6.6} are expressions
\begin{equation}\label{6.7}
\bigl\{\nabla_{\eta^\mu}^\pm , \nabla_{\eta^\nb}^\pm \bigr\}=\pm\im\eta_{\mu\nb}^{}
\end{equation}
with an imaginary unit on the right-hand side.

\medskip

\noindent{\bf Grassmann variables $\tilde\eta^\mu , \tilde\eta^\mb$.} We introduced connections \eqref{6.5} and covariant derivatives \eqref{6.6} following  \cite{Kost}. However, operators \eqref{6.6} in matrix representation are  identified with gamma matrices, for anticommutators of which on the right-hand side of \eqref{6.7} one usually uses $\eta_{\mu\nb}^{}$ without an imaginary unit. To move to such operators, one should redefine Grassmann variables by introducing 
\begin{equation}\label{6.8}
\tilde\eta^\mu =e^{\im\frac{\pi}4}\eta^\mu =\frac{1}{\sqrt 2}\bigl(\tilde\xi^\mu + \im\tilde\xi^{\mu+4}\bigr)\und
\tilde\eta^\mb =e^{\im\frac{\pi}4}\eta^\mb =\frac{1}{\sqrt 2}\bigl(\tilde\xi^\mu - \im\tilde\xi^{\mu+4}\bigr)
\end{equation}
with $\tilde\xi^M=e^{\im\frac{\pi}4}\xi^M$. At the same time the reality properties of the original Grassmann variables should be changed by assuming that  $(\xi^M)^*=\im\xi^M$ (Rodgers convention). Then for new variables $(\tilde\xi^M)^*=\tilde\xi^M$ will be hold (DeWitt convention) so that $(\tilde\eta^\mu)^*=\tilde\eta^\mb$. In these variables, the covariant derivatives \eqref{6.6} take the form
\begin{equation}\label{6.9}
\nabla^\pm_{\tilde\eta^\mu}=\dpar_{\tilde\eta^\mu}^{}\pm\sfrac12\eta_{\mu\nb}^{}\tilde\eta^\nb\und
\nabla^\pm_{\tilde\eta^\mb}=\dpar_{\tilde\eta^\mb}^{}\pm\sfrac12\eta_{\mb\nu}^{}\tilde\eta^\nu
\end{equation}
and the anticommutators \eqref{6.7} are replaced by the following
\begin{equation}\label{6.10}
\bigl\{\nabla^\pm_{\tilde\eta^\mu}, \nabla^\pm_{\tilde\eta^\nb}\bigr\}=\pm\eta_{\mu\nb}^{}\ .
\end{equation}
Note that the right-hand sides in  \eqref{6.7} and  \eqref{6.10} are components of the curvature tensor $F_{\sf vac}^\pm$ of the connection \eqref{6.3} along the fermionic directions.

\medskip

\noindent{\bf Grassmann variables $\eta^\mu_\pm$.} The operators $\nabla^+_{\tilde\eta^\mu}$ and $\nabla^+_{\tilde\eta^\mb}$ in the canonical anticommutation relations (CAR) of the form \eqref{6.10} can be identified with the fermionic annihilation and creation operators $b_{\mu +}^{}$ and $b_{\mb +}^{\+}$. For operators $\nabla^-_{\tilde\eta^\mu}$ and $\nabla^-_{\tilde\eta^\mb}$ this is not the case due to the minus on the right side. This can be corrected by introducing variables
\begin{equation}\label{6.11}
\eta_+^\mu :=\tilde\eta^\mu\ ,\quad \eta_+^\mb :=\tilde\eta^\mb\und
\eta_-^\mu :=-\im\tilde\eta^\mb\ ,\quad \eta_-^\mb :=-\im\tilde\eta^\mu\ ,
\end{equation}
where $\eta_-^\mu$ replaces $\tilde\eta^\mb$. In this variables the covariant derivatives \eqref{6.9} take on the same form
\begin{equation}\label{6.12}
\nabla^{}_{\eta_\pm^\mu}=\dpar^{}_{\eta_\pm^\mu}+\sfrac12\,\eta_{\mu\nb}^{}\eta_\pm^\nb\und 
\nabla^{}_{\eta_\pm^\mb}=\dpar^{}_{\eta_\pm^\mb}+\sfrac12\,\eta_{\mb\nu}^{}\eta_\pm^\nu
\end{equation}
and we obtain
\begin{equation}\label{6.13}
\bigl\{\nabla^{}_{\eta^\mu_\pm}, \nabla^{}_{\eta^\nb_\pm}\bigr\}=\eta_{\mu\nb}^{}\ .
\end{equation}
Now we can put $b_{\mu\pm}^{}=\nabla^{}_{\eta^\mu_\pm}$ and $b_{\mb\pm}^{\+}=\nabla^{}_{\eta^\mb_\pm}$. Note that $\nabla^{}_{\eta^\mb_-}$ defines holomorphic objects from the point of view of the complex structure $\CJ_-=-\CJ$ and antiholomorphic objects from the point of view of the complex structure $\CJ_+=\CJ$. 

\medskip

\noindent{\bf Polarizations and CAR.} We introduced CAR for operators \eqref{6.12} acting on Grassmann algebrae $\Lambda (\R^8)\otimes V^\pm\cong\Lambda (\C^8)$. The representation of the algebra  \eqref{6.13} on these spaces is not irreducible, just as it was for the CCR in the bosonic case. Recall that $\C^8 =\C^4\oplus\bar\C^4$ and an irreducible representation of CAR can be defined on the space $S^+=\Lambda(\C^4)\subset\Lambda(\C^4\oplus\bar\C^4)$. Elements of $S^+$ are spinors. Similarly, one can define a representation of CAR on the space $S^-=\Lambda(\bar\C^4)\subset\Lambda(\C^4\oplus\bar\C^4)$ of charge-conjugate spinors. The Dirac equation is defined for $S=S^+\oplus S^-$, where $S^+$ parametrizes particles and $S^-$ parametrizes antiparticles.

To introduce subspaces $S^+$ and $S^-$ in the space $\Lambda(\C^8)$ we define the fermionic Dolbeault operators
\begin{equation}\label{6.14}
\bar\dpar_\pm^{\,\tt F}=\dd\eta_\pm^{\mb}\bigl(\dpar_{\eta_\pm^\mb}^{} - \sfrac12\,\eta^{}_{\mb\nu}\eta_\pm^\nu\bigr)
\end{equation}
and impose on $\Psi_\pm^{}\in\Lambda(\C^8)$ the polarization conditions
\begin{equation}\label{6.15}
\bar\dpar_\pm^{\,\tt F}\Psi_\pm^{}=0\ .
\end{equation}
Their solutions are
\begin{equation}\label{6.16}
\Psi_\pm^{}=\psi_\pm^{} (z_\pm^{}, \eta_\pm^{}, \tau)\,\psi_0^{\tt B}\,\psi_0^{\tt F}v_\pm^{}\ ,
\end{equation}
where $\Psi_\pm^{}$ are chosen to satisfy the bosonic polarization conditions \eqref{5.8}. Here the function $\psi_0^{\tt F}$ (fermionic ground state) has the form
\begin{equation}\label{6.17}
\psi_0^{\tt F}=\exp(-\sfrac12\eta^{}_{\mu\nb}\tilde\eta^\mu\tilde\eta^\nb)=
\exp(-\sfrac12\eta^{}_{\mu\nb}\eta^\mu_+\eta^\nb_+)=
\exp(-\sfrac12\eta^{}_{\mu\nb}\eta^\mu_-\eta^\nb_-)\ .
\end{equation}
Thus, we obtain spaces $\CF^\pm=\CF^\pm_{\tt B}\otimes S^\pm$ of polarized sections $\Psi_\pm^{}$ of sheaves $\CL_\C^\pm$ of the form \eqref{6.16}, where $\psi_+$ and $\psi_-$ are holomorphic and antiholomorphic functions of $z^\mu , \tilde\eta^\mu$ on the superspace \eqref{4.12} defining the supersymmetric classical Klein-Gordon oscillator.

In addition to formulae \eqref{5.12} for the action on $\Psi_\pm$ from \eqref{6.16} by the bosonic creation and annihilation operators, we also have formulae for the action of the fermionic creation and annihilation operators:
\begin{equation}\label{6.18}
\nabla^{}_{\eta_\pm^\mu}\Psi_\pm^{}=(\dpar^{}_{\eta_\pm^\mu}\psi_\pm^{})\,\psi_0^{\tt F}\,v_\pm^c
\und 
\nabla^{}_{\eta_\pm^\mb}\Psi_\pm^{}=(\eta_{\mb\nu}^{}\eta_\pm^\nu\psi_\pm^{})\,\psi_0^{\tt F}\,v_\pm^c\ .
\end{equation}
From \eqref{5.9}-\eqref{5.13} and  \eqref{6.16}-\eqref{6.18} it follows that we have irreducible representation of CCR \eqref{5.13} and CAR \eqref{6.13} on the Fock spaces $\CF^\pm =\CF^\pm_{\tt B}\otimes S^\pm$ of holomorphic functions of the form  \eqref{6.16}.

\medskip

\noindent{\bf Inner products on $S^\pm$.} Functions $\Psi_+$ and $\Psi_-$ from \eqref{6.16} can be expanded in Grassmann variables $\eta_+^\mu$ and $\eta_-^\mu$, respectively. The coefficients of the powers of $\eta_\pm^\mu$ will be holomorphic functions on the spaces $\C_\pm^{3,1}=(\R^{6,2}, \pm\CJ )$ with values in the spaces of spinors $S^\pm =\Lambda (\C^{3,1}_\pm)$. On the spaces $S^\pm$ we introduce the standard inner product
\begin{equation}\label{6.19}
(\Psi_\pm^{}, \Psi_\pm^{})=\int_{\Pi\C_\pm^{3,1}}\Psi_\pm^{\+} \Psi_\pm^{}\dd V_{\tt F} ^{}=
(\psi_0^{\tt B})^2\int_{\Pi\C_\pm^{3,1}}\psi_\pm^*\psi_\pm^{}(\psi_0^{\tt F})^2\dd V_{\tt F} ^{}\ ,
\end{equation}
where $\dd V_{\tt F} ^{}=\prod^3_{\mu =0}\dd\tilde\eta^\mu\tilde\eta^\mb, (\psi_0^{\tt B})^2=e^N$ and the square $(\psi_0^{\tt F})^2$ of the function \eqref{6.17} defines the fermionic measure.

To calculate the integrals \eqref{6.19} we expand $\psi_\pm^{}$ in $\eta_\pm^\mu$ and drop ``$\pm$" indices in formulae, since formulae are the same for $\psi_+$ and $\psi_-$. We have $\psi=\varrho +\eta^0\vsig$, where
\begin{equation}\label{6.20}
\begin{split}
\vrho &=\vrho_e + \vrho_a\eta^a +\sfrac12\tilde\vrho_a\veps_{bc}^a\eta^b\eta^c+\tilde\vrho_e\eta^1\eta^2\eta^3\with \tilde\vrho_a=\sfrac12\veps_{abc}\vrho_{bc},\ \tilde\vrho_e=\vrho_{123}\ ,\\[2pt]
\vsig &=\vsig_{0e}^{}+ \vsig_{0a}^{}\eta^a +\sfrac12\tilde\vsig_{0a}\veps_{bc}^a\eta^b\eta^c+\tilde\vsig^{}_{0e}\eta^1\eta^2\eta^3\with \tilde\vsig^{}_{0a}=\sfrac12\veps_{abc}\vsig^{}_{0bc},\ \tilde\vsig^{}_{0e}=\vsig^{}_{0123}\ .
\end{split}
\end{equation}
Substituting expansion \eqref{6.20} into \eqref{6.19} and integrating over $\tilde\eta^\mu$, we obtain
\begin{equation}\label{6.21}
\begin{split}
e^{-N}(\Psi , \Psi)&=(\vrho , \vrho)-(\vsig , \vsig)=(\vrho_{\bar e}^*\vrho_e + \delta^{\ab b}\vrho_{\ab}^*\vrho_b+
+ \delta^{\ab b}\tilde\vrho_{\ab}^*\tilde\vrho_b+\tilde\vrho_{\bar e}^*\tilde\vrho_e)
\\[2pt]
&-(\vsig^*_{\bar 0\bar e}\vsig_{0e}+ \delta^{\ab b}\vsig^*_{\bar 0\bar a}\vsig_{0b}+ \delta^{\ab b}\tilde\vsig^*_{\bar 0\bar a}\tilde\vsig_{0b}+\tilde\vsig^*_{\bar 0\bar e}\tilde\vsig_{0e})\  ,
\end{split}
\end{equation}
where $(\vrho , \vrho )$ and $(\vsig , \vsig )$ is the inner producton the space $\Lambda (\C^3)$.

Recall that the bosonic part $Z_6$ of the covariant phase space $\Pi TZ_6$ is a K\"ahler-Einstein manifold and the metric on it is positive definite. Moreover, the tangent bundle to $Z_6$ is trivializable, $TZ_6\cong Z_6\times \R^6\cong Z_6\times\C^3$, since $Z_6$ is covered by one chart. Let us also recall that $B_\pm^3=(Z_6, \pm\CJ)\subset\C_\pm^{3,1}$ and therefore $\eta_\pm^a$ are Grassmann variables on the tangent spaces $\C_\pm^3$ to $B_\pm^3$, and $\eta_\pm^0$ are variables on the one-dimensional complex subspaces in $\C_\pm^{3,1}$ orthogonal to $B_\pm^3$. In the expansion $\psi_\pm =\vrho_\pm +\eta_\pm^0\vsig_\pm$ we have functions $\vrho_\pm$ of $\eta_\pm^a$ which are free spinors on $B_\pm^3$, and the non-physical parts $\vsig_\pm$ of the spinors $\psi_\pm$ appear when lifting $\vrho_\pm$ from $B_\pm^3$ to $\C_\pm^{3,1}$. Below we will show that $\vsig_\pm$ are expressed in terms of free spinors $\vrho_\pm$ in the general solution of the Dirac equation on the phase space $T^*\R^{3,1}$.

\medskip

\noindent{\bf Dirac operators $\Gamma_\pm$.} In \eqref{5.9}-\eqref{5.13} and \eqref{6.16}-\eqref{6.18} we introduced  Fock spaces  $\CF^\pm{=}\CF^\pm_{\tt B}{\otimes}S^\pm$ of holomorphic and antiholomorphic functions on the space $\R^8\times\Pi\R^8$. Thus we have a direct sum of spaces,
\begin{equation}\label{6.22}
\CF =\CF^+\oplus\CF^-=\CF^+_{\tt B}\otimes S^+\oplus\CF^-_{\tt B}\otimes S^-\subset \CL_\C^+\oplus\CL_\C^-=\CL_{\C^2}^{}\ .
\end{equation}
On the space \eqref{6.22} act bosonic covariant derivative \eqref{5.16} and fermionic covariant derivatives
\begin{equation}\label{6.23}
\nabla_\mu^{\tt F} =v_+^{}v_+^\+\nabla_{\eta_+^\mu}^{}+v_-^{}v_-^\+\nabla_{\eta_-^\mu}^{}=
\sfrac12 (1+Q_{\sf v})\nabla_{\eta_+^\mu}^{}+\sfrac12 (1-Q_{\sf v})\nabla_{\eta_-^\mu}^{}\ ,
\end{equation}
where $Q_{\sf v}=-\im J$ is the quantum charge operator. When acting on $\Psi_\pm^{}$, the operator $\nabla_\mu^{\tt F}$ coincides with $\nabla_{\eta_\pm^\mu}^{}$.

For the bosonic part $\mu^{\tt B}_{H_0}$ of the momentum map \eqref{3.8} we introduced the covariant Laplace operator \eqref{5.15}, which reduces to operators \eqref{5.23} when acting on spaces $\CF^\pm_{\tt B}$. For the fermionic momentum map $\mu_{H_1}^{}$ from  \eqref{3.18} defining an odd tangent bundle $\Pi T\AdS_7$, the operator version is the covariant odd Laplacian
\begin{equation}\label{6.24}
\hat\mu_{H_1}^{}=\sqrt 2\,\eta^{\mu\nb}(\nabla_\mu^{\tt B}\nabla_\nb^{\tt F}+\nabla_\mu^{\tt F}\nabla_\nb^{\tt B})\ ,
\end{equation}
where the bosonic and fermionic covariant derivatives $\nabla^{\tt B}$ and $\nabla^{\tt F}$ are given in \eqref{5.16} and \eqref{6.23}, respectively. When acting on $\Psi =\Psi_+\oplus\Psi_-\in\CF$ from \eqref{6.22}, this operator splits into a direct sum of operators,
\begin{equation}\label{6.25}
\hat\mu_{H_1}^{}=\hat\mu_{H_1}^{+}\oplus\hat\mu_{H_1}^{-}\with
\hat\mu_{H_1}^{\pm}=\sqrt 2\,\eta^{\mu\nb}(\nabla_{z^\mu_\pm}^{}\nabla_{\eta^\nb_\pm}^{}+\nabla_{\eta^\mu_\pm}^{}\nabla_{z^\nb_\pm}^{})\ ,
\end{equation}
since the covariant derivatives \eqref{6.23} have a different form when acting on $\Psi_+$ and $\Psi_-$ with $\qv=1$ and $\qv=-1$. It is not difficult to verify that for $\Psi_\pm^{}$ from \eqref{6.16} we have
\begin{equation}\label{6.26}
\hat\mu_{H_1}^{}\Psi=\hat\mu_{H_1}^{+}\Psi_+^{} + \hat\mu_{H_1}^{-}\Psi_-^{} =
\sqrt 2\, (\Gamma_+\psi_+ v_+ + \Gamma_-\psi_- v_-)\,\psi_0^{\tt B}\,\psi_0^{\tt F}\ ,
\end{equation}
where the operators $\Gamma_\pm$ are vector fields
\begin{equation}\label{6.27}
\Gamma_\pm =\eta_\pm^\mu\frac{\dpar}{\dpar z_\pm^\mu} - \frac{z_\pm^\mu}{w^2}\frac{\dpar}{\dpar \eta_\pm^\mu}
\end{equation}
defined on superspaces $\C_\pm^{3,1}\times \Pi\C_\pm^{3,1}$. The operators \eqref{6.27} define the Dirac equation on the phase space. They will take a more familiar matrix form for the components in the expansion of $\psi_\pm$ in Grassmann variables $\eta_\pm^\mu$.

{\bf Remark.} To better understand the geometric meaning of the Dirac operators \eqref{6.27}, let us rewrite the Hamiltonian vector field $V^{}_{H_1}$ from \eqref{3.21} in terms of coordinates $z_\pm^\mu$ and $\eta_\pm^\mu$. We obtain
\begin{equation}\label{6.28}
V^{}_{H_1}=e^{\im\frac{\pi}{4}}\,\Gamma_++e^{-\im\frac{\pi}{4}}\,\Gamma_-\ ,
\end{equation}
where $\Gamma_+$ is a holomorphic and  $\Gamma_-$ is an antiholomorphic vector field on $\R^8\times\Pi\R^8$. These vector fields together with the vector field $V^{}_{H_0}$ are the generators \eqref{4.2} of the supergroup acting on the level set $\Pi T\AdS_7$ on which the oscillating spinning particles are located. Factorization by this group defines the spaces $\Pi T B^3_\pm$ of initial data for particles and antiparticles. For quantum particles, these vector fields define field equations projecting the Fock space $\CF$ from \eqref{6.22} onto the direct sum of weighted Bergman spaces of functions on covariant phase spaces $B_\pm^3\subset\Pi T B^3_\pm$. For free particles the description is similar, we will discuss it elsewhere.

\medskip

\noindent{\bf Evolution equation.} The spinor field $\Psi_+$ and $\Psi_-$ belong to conjugate spaces $\CL_\C^+$ and $\CL_\C^-$, so the equations defining their evolution are conjugate. We define them in the form
\begin{equation}\label{6.29}
J\dpar_\tau\Psi =\hat\mu_{H_1}^{}\Psi\ \ \Rightarrow\ \ \pm\im\dpar_\tau\Psi = \hat\mu_{H_1}^{\pm}\Psi^{}_{\pm}\for \Psi =\Psi_+ +\Psi_-\ .
\end{equation}
Substituting in \eqref{6.29} the explicit form of $\Psi^{}_{\pm}$ from \eqref{6.16} with a dependence on $\tau$ of the
form $\exp(\mp\im m\tau)$, we obtain the equations
\begin{equation}\label{6.30}
\bigl(\eta_\pm^\mu\frac{\dpar}{\dpar z_\pm^\mu} - \frac{z_\pm^\mu}{w^2}\frac{\dpar}{\dpar \eta_\pm^\mu}-\frac{m}{\sqrt 2}\bigr)\,\psi_\pm^{}(z_\pm^{} , \eta_\pm^{})=0\ .
\end{equation}
These are the Dirac equations on the phase space $T^*\R^{3,1}$ for holomorphic fields $\Psi_+\in\CL_\C^+$ (particles) and antiholomorphic fields $\Psi_-\in\CL_\C^-$ (antiparticles). By acting on \eqref{6.30} with the operator $\Gamma_\pm -m/\sqrt 2$ in parentheses, we obtain the equation of the Klein-Gordon oscillator of the form
\begin{equation}\label{6.31}
\bigl(z_\pm^\mu\frac{\dpar}{\dpar z_\pm^\mu}+\eta_\pm^\mu\frac{\dpar}{\dpar \eta_\pm^\mu}+N\bigr)\,\psi_\pm =0\ ,
\end{equation}
where the number $N=m/2\omega$ was introduced in \eqref{5.24}. Equations \eqref{6.31} differ from equations \eqref{5.23} by the term $\eta_\pm^\mu\,{\dpar}/{\dpar \eta_\pm^\mu}$ arising from the dependence of $\psi_\pm$ on the Grassmann variables $\eta_\pm^\mu$. They are a quantum version of the momentum map function \eqref{3.8} splitting into holomorphic and antiholomorphic parts of the vector field $V_{H_0}^{}$ from \eqref{3.4}.

\medskip

\noindent{\bf Solutions.} To solve the Dirac equations \eqref{6.30}, we write $\psi_\pm$ in the form
\begin{equation}\label{6.32}
\psi_\pm^{}=\vrho_\pm^{}+\eta_\pm^{0}\vsig_\pm^{}\ ,
\end{equation}
where the functions $\vrho_\pm^{}$ and $\vsig_\pm^{}$ do not depend on $\eta_\pm^{0}$. Substituting \eqref{6.32} into \eqref{6.30} we obtain the equations
\begin{equation}\label{6.33}
\vsig_\pm^{}=\frac{w^2}{z_\pm^0} \Bigl(\Upsilon_\pm - \frac{m}{\sqrt 2} \Bigr)\,\vrho_\pm^{}\for
\Upsilon_\pm :=\eta_\pm^a \frac{\dpar}{\dpar z_\pm^a} - \frac{z_\pm^a}{w^2} \frac{\dpar}{\dpar \eta_\pm^a}   
\end{equation}
and
\begin{equation}\label{6.34}
\frac{\dpar\vrho_\pm^{}}{\dpar z_\pm^0}-\Bigl(\Upsilon_\pm + \frac{m}{\sqrt 2} \Bigr)\,\vsig_\pm^{}=0\ \ \Rightarrow\ \
\Bigl(z_\pm^\mu \dpar_{z_\pm^\mu}^{} +\eta_\pm^\mu \dpar_{\eta_\pm^\mu}^{}+N\Bigr)\,\vrho_\pm^{}=0\ .
\end{equation}
Equations  \eqref{6.34} for $\vrho_\pm^{}$ coincide with the Klein-Gordon oscillator equation with $\vrho_\pm^{}$ independent of $\eta_\pm^0$.

After using the expansion \eqref{6.20}, equations \eqref{6.34} break down into equations
\begin{equation}\label{6.35}
\begin{split}
z_\pm^\mu\dpar_{z_\pm^\mu}^{}\,\vrho^\pm_{e}+N\vrho^\pm_{e}=0\ \ &\Rightarrow\ \
\vrho^\pm_{e}=\Bigl(\frac{1}{\sqrt 2\omega z_\pm^0}\Bigr)^N\chi_e^\pm (y_\pm )\ ,\\[2pt]
z_\pm^\mu\dpar_{z_\pm^\mu}^{}\,\vrho^\pm_{a}+(N+1)\vrho^\pm_{a}=0\ \ &\Rightarrow\ \
\vrho^\pm_{a}=\Bigl(\frac{1}{\sqrt 2\omega z_\pm^0}\Bigr)^{N+1}\chi_a^\pm (y_\pm )\ ,\\[2pt]
z_\pm^\mu\dpar_{z_\pm^\mu}^{}\,\tilde\vrho^\pm_{a}+(N+2)\tilde\vrho^\pm_{a}=0\ \ &\Rightarrow\ \
\tilde\vrho^\pm_{a}=\Bigl(\frac{1}{\sqrt 2\omega z_\pm^0}\Bigr)^{N+2}\tilde\chi_a^\pm (y_\pm )\ ,\\[2pt]
z_\pm^\mu\dpar_{z_\pm^\mu}^{}\,\tilde\vrho^\pm_{e}+(N+3)\tilde\vrho^\pm_{e}=0\ \ &\Rightarrow\ \
\tilde\vrho^\pm_{e}=\Bigl(\frac{1}{\sqrt 2\omega z_\pm^0}\Bigr)^{N+3}\tilde\chi_e^\pm (y_\pm )\ ,
\end{split}
\end{equation}
where $y_+^a\in B_+^3=(Z_6, \CJ)$ and $y_-^a=\overline{y_+^a}\in B_-^3=(Z_6, -\CJ)$. Recall that indices ``$\pm$"  denote holomorphicity on spaces $B_\pm^3$. Equivalently, we can speak of holomorphic functions (particles) and antiholomorphic functions (antiparticles) on the same manifold $H_\C^3=B_+^3=(Z_6, \CJ)$ and bundles  $\Pi T^{1,0}H_\C^3$ and  $\Pi T^{0,1}H_\C^3$ of type (1,0) and (0,1) over it.

Thus, the general solution of the Dirac oscillator equation on the phase space has the form
\begin{equation}\label{6.36}
\Psi =\Bigl(e^{-\im m\tau}\psi_+(z_+, \eta_+)v_+ +e^{\im m\tau}\psi_-(z_-, \eta_-)v_-\Bigr)\,\psi_0^{\tt B}\,\psi_0^{\tt F}\ ,
\end{equation}
where $\psi_\pm$ have the form \eqref{6.32}, the functions $\vrho_\pm$ and $\vsig_\pm$ are polynomials in $\eta_\pm^a$ of the form \eqref{6.20}, $\vsig_\pm$ are expressed through $\vrho_\pm$ by formula \eqref{6.33}, and the explicit form of the eight functions $(\vrho_e^+, \vrho_a^+, \tilde\vrho_a^+, \tilde\vrho_e^+)$ (particles) and eight functions $(\vrho_e^-, \vrho_a^-, \tilde\vrho_a^-, \tilde\vrho_e^-)$ (antiparticles) is given in \eqref{6.35}.
These functions belong to eight Bergman spaces,
\begin{equation}\label{6.37}
L_h^2(B_+^3, \mu_{N+k}^{})\und L_h^2(B_-^3, \mu_{N+k}^{})\ ,
\end{equation}
with measures $\mu_{N+k}^{}$, $k=0,1,2,3$. The inner product is given by formulae \eqref{6.19}-\eqref{6.21} and it is positive definite as the norm for spinors on six-dimensional K\"ahler-Einstein manifolds $(Z_6, \pm\CJ)$.

To summarize, we have shown that the general solution of the supersymmetric Klein-Gordon oscillator equations in the complex Bargmann-Fock-Segal representation is a direct sum of solutions $\Psi_+$ and $\Psi_-$ parametrized by the Bergman spaces \eqref{6.37}. Thus, this supersymmetric model is exactly solvable, Lorentz covariant and unitary.

\bigskip

\noindent 
{\bf\large Acknowledgments}

\noindent
I am grateful to Tatiana Ivanova for useful remarks.


\end{document}